\documentclass[singlecolumn]{aastex6}
\usepackage{amsmath}
\newcommand{\fluxunits}{erg cm$^{-2}$ s$^{-1}$}
\newcommand{\cgs}{erg cm$^{-2}$ s$^{-1}$ \AA$^{-1}$ sr$^{-1}$}
\newcommand{\dmehard}{F13$\_$hard$\_$dMe}
\newcommand{\dmesoft}{F13$\_$soft$\_$dMe}
\newcommand{\solsoft}{5F11$\_$soft$\_$sol}
\newcommand{\solhard}{F13$\_$hard$\_$sol}

\shorttitle{A simple flare compression model}
\shortauthors{Kowalski et al.\ }

\begin{document}

\title{Parameterizations of Chromospheric Condensations in \MakeLowercase{d}G and \MakeLowercase{d}M\MakeLowercase{e} Model Flare Atmospheres}
\author{Adam F. Kowalski}
\affil{Department of Astrophysical and Planetary Sciences, University of Colorado Boulder, 2000 Colorado Ave, Boulder, CO 80305, USA.}
\affil{National Solar Observatory, University of Colorado Boulder, 3665 Discovery Drive, Boulder, CO 80303, USA.}
\affil{Laboratory for Atmospheric and Space Physics, University of Colorado Boulder, 3665 Discovery Drive, Boulder, CO 80303, USA.}
\email{adam.f.kowalski@colorado.edu}
\author{Joel C. Allred}
\affil{NASA/Goddard Space Flight Center, Code 671, Greenbelt, MD 20771}
\begin{abstract}
The origin of the near-ultraviolet and optical continuum
radiation in flares is critical for understanding
particle acceleration and impulsive heating in stellar
atmospheres.  Radiative-hydrodynamic simulations in 1D have 
shown that high energy deposition rates from electron beams produce two
flaring
layers at $T\sim10^4$ K that develop in the chromosphere:  a cooling 
condensation (downflowing compression) and heated non-moving (stationary) flare layers
just below the condensation.  These atmospheres reproduce several
observed phenomena in flare spectra, such
as the red wing asymmetry of the emission lines in solar flares and a 
small Balmer jump ratio in M dwarf flares.  The
high beam flux simulations are
computationally expensive in 1D, and the (human) timescales 
for completing NLTE models with adaptive grids in 3D will likely be unwieldy for a time
to come.  We have developed a prescription for predicting
the approximate evolved states, continuum optical depth, and the emergent continuum
flux spectra of radiative-hydrodynamic model flare atmospheres.  These
approximate prescriptions are based on an important atmospheric parameter:  the column mass ($m_{\rm{ref}}$) at which hydrogen
becomes nearly completely ionized at the depths that are approximately in
steady state with the electron beam heating.  Using this new modeling
approach, we find that high energy flux density ($>$F11)
electron beams are needed to reproduce the brightest observed continuum intensity in IRIS
data of the 2014-Mar-29 X1 solar flare and that variation in
$m_{\rm{ref}}$ from 0.001 to 0.02 g cm$^{-2}$ reproduces most of the observed
range of the optical continuum flux ratios at the peaks of M dwarf flares.  

\end{abstract}

\section{Introduction}

Stellar flares are thought to be produced from the atmospheric heating that results from coronal magnetic field reconnection
and retraction.  Ambient electrons, protons, and
heavy nuclei are accelerated to very high energies and produce the observed X-ray and
gamma-ray emissions. 
The hard X-ray emission at tens to hundreds of keV is cospatial and
cotemperal with typical signatures of chromospheric heating, such as
H$\alpha$ ribbons and the near-ultraviolet (NUV;
  $\lambda=2000-4000$ \AA) and optical ($\lambda=4000-7000$ \AA) continuum
radiation.  Thus, these high energy particles are
likely the source of powering most of the chromospheric heating and radiative
response.  Moreover, recent high spatial
resolution imagery of NUV and optical footpoints in solar flares suggest a very high electron beam
flux density \citep{Fletcher2007, Krucker2011, Kleint2016, Jing2016,
  Kowalski2017A, Sharykin2017} which may be difficult to sustain due to plasma
instabilities \citep{Lee2008, Li2014}.

Radiative-hydrodynamic (RHD) simulations with heating from high beam flux
densities nonetheless provide important insights into the atmospheric response that produces
 several well-observed spectral phenomena in M dwarf and solar flares.  
A high flux density
 of nonthermal electrons with a low energy cutoff of $E_c = 20-40$ keV produces dense, downflowing
 compressions that originate in the mid to upper chromosphere.  These chromospheric
 compressions (or chromospheric condensations, hereafter ``CC"s) 
 have physical depth ranges in the atmosphere of $\Delta z \sim20-30$ km.  The CCs evolve from high temperature and
low density to low temperature ($T\sim$10,000 K) and much higher
density as they compress and descend to the lower
chromosphere \citep{Kennedy2015, Kowalski2015}; most of the CC evolution occurs on timescales of seconds to several
tens of seconds \citep{Fisher1989};  CCs have been extensively modeled 
in the literature \citep{Livshits1981, Emslie1985, Fisher1985a,
  Fisher1985b, Canfield1987, Gan1992}. 
  
 In solar flares, compelling observational evidence exists for the formation of these CCs.  The spectrally
resolved red-wing
asymmetry (often referred to as ``RWA'') in chromospheric lines such as H$\alpha$, Mg II, and Fe II is frequently observed in the
impulsive phase of flares \citep{Ichimoto1984, Graham2015,
  Kowalski2017A}.  These RWAs exhibit spectrally resolved peaks with redshifts of 
$\lambda-\lambda_{\rm{rest}}=15-140$ km s$^{-1}$.  The
brightness of the RWA in NUV Fe II lines relative to
the intensity of the line component at the rest wavelength has been
reproduced with a high flux electron beam of 5x10$^{11}$ \fluxunits\ \citep[hereafter, 5F11;][]{Kowalski2017A}.  

In magnetically active M dwarf (dMe) flares, the observed NUV and
optical flare continuum (sometimes referred to as white-light radiation if
detected in broadband optical radiation on the Sun or in the Johnson
$U$-band in dMe stars) distribution can
be 
reproduced in 1D model snapshots of a very dense, evolved CC that results from the extremely high
energy flux density
($\sim 10^{13}$ \fluxunits, hereafter F13) in nonthermal electron beams
 lasting several seconds \citep{Kowalski2015, Kowalski2016,
   Kowalski2017B}.  This flux density is expected to result in beam
instabilities (even in the much larger ambient coronal densities of
dMe stars) and a strong return current electric field
\citep[e.g.,][]{VandenOord1990}.  Also, the hydrogen Balmer line broadening predicted from
these evolved CCs far exceeds
the typical values that are observed without including several other lower density
emitting regions in the modeling \citep{Kowalski2017B}.  Alternative heating scenarios may be necessary to reproduce the 
continuum radiation, such as a very high low-energy cutoff
\citep{Kowalski2017B}, high energy proton/ion beams, or possibly
Alfven wave heating \citep{Reep2016, Kerr2016}.   Only a
limited range of heating simulations with electron beam
flux densities between $\sim 10^{12}$ and $\sim 10^{13}$ \fluxunits\ has been
tested to
determine if such a high beam density of  $\sim 10^{13}$ \fluxunits\ is required to produce
the observed range of flare continuum flux ratios and Balmer line
broadening in dMe flares. 

The atmospheric response to the energy deposition from a high beam
flux density of nonthermal electrons results in complete helium
ionization and a thermal instability as the temperature exceeds the
peak of the radiative loss function for C and O ions at $T\sim100,000
- 200,000$ K.  The chromospheric temperature initially at $T\sim7000$
K exceeds 10 MK in less than a fraction of a second after beam heating
begins; this localized explosion in the chromosphere results in large
temperature, density, pressure, and ionization fraction changes over very
narrow height ranges, $\Delta z \sim 10-15$ meters.  One-dimensional
RHD simulations have the advantage of resolving these gradients 
with an adaptive grid
\citep{Dorfi1987}, but the atmospheric evolution can take weeks to several months
to compute \citep{Abbett1999, Allred2005, Allred2006,
  Kowalski2015, Kennedy2015}.  The small time steps ($10^{-7}-10^{-8}$~s) in these calculations are caused by 
the accuracy of the helium population convergence at these steep gradients and are exacerbated by a radiative instability
in a very narrow ($\sim5$ km), cool region
between the flare corona and the large temperature
gradient at the chromospheric explosion \citep[see][]{Kennedy2015}.  The onset threshold of explosive evaporation
and condensation in the chromosphere
depends upon all of the parameters that characterize electron beam heating 
\citep{Fisher1989} but
generally occurs at high energy flux densities, typically
exceeding $10^{11}$ \fluxunits\ (F11) with a moderate low-energy
cutoff $E_{\rm{c}}$ value \citep[$E_{\rm{c}} \sim25$ keV;][]{Kowalski2017A}.

There are far less constraints on the electron beam parameters in dMe flares because the
hard X-ray flux is faint except during extreme events;  in these events, 
the hard X-ray emission can be explained by a superhot thermal
component \citep{Osten2007, Osten2010, Osten2016}.   Radio 
observations directly probe mildly relativistic electrons in dMe flares, but one must
observe at optically thin frequencies in order to relate the power law
index of the radio emission directly to the power law index of electrons
\citep{Dulk1985, Osten2005, Smith2005, Osten2016}.   Due to the large
contrast at NUV and blue optical wavelengths, flare spectra around the
Balmer limit wavelength are the most direct way to probe the impulsive
release of magnetic energy in dMe flares.  
Models over a large parameter space of electron beam heating can then be
used to infer the properties of the accelerated particles in these
very active stars.  Large grids of models would also show if very 
high beam flux densities ($\sim$F13) are required to produce the observed spectral
properties in dMe flares.  This would motivate improving the treatment
of electron beam propagation to include the effects of the return current
electric field and plasma instabilities.

Modeling the full
RHD response even in 1D for a large parameter space of electron
beam distributions with very high heating rates is currently
computationally challenging and will become more time-consuming when
3D models that employ adaptive grids for resolving steep pressure
gradients are developed in the future. 
In this paper, we present a method for obtaining prompt insight into the 
evolution of the radiative and hydrodynamic response to high energy
deposition rates from electron beams with a low-to-moderate low-energy
cutoff ($E_c=15-40$ keV), thus providing important guidance about
which heating models are most interesting to follow with RADYN for the full evolution.
This paper is organized as follows:  In section 2.1 we summarize the
response to high electron beam flux densities in solar and dMe atmospheres
that are used in the analysis, in section 2.2 we describe our
analysis and prescription for approximating the RHD models of the
evolved states of these flare atmospheres, in section 3
we discuss several applications for our approximate model atmospheres,
and in section 4 we present several new conclusions about flares that
are based off of our
approximations.  In an appendix, we show that our new modeling
prescriptions can be used to produce broad wavelength flare 
spectral predictions.

\section{Approximations to Dynamic Flare Atmospheres}\label{sec:method}

\subsection{High Beam Flux Density RHD Modeling with the RADYN Code}

High electron beam flux density simulations with a low-energy cutoff of $E_c=20-40$ keV produce two dense layers at low temperature ($T\lesssim 13,000$ K) at pre-flare chromospheric heights which flare brightly in NUV and optical radiation \citep{Kowalski2015, Kowalski2016, Kowalski2016IAU, Kowalski2017A}.  The electron beam distribution is characterized by a power-law index, and most of the beam energy is thus concentrated near the low-energy cutoff.  The two flaring layers that develop from high energy deposition rates in the mid to upper chromosphere are the following:

\begin{enumerate}
\item  A downflowing ($v\sim50-100$ km s$^{-1}$), hot ($T\sim10,000-13,000$ K) and dense (several x $10^{14}$ to several x$10^{15}$ cm$^{-3}$) region
that is several tens of km in vertical extent just below a lower steep pressure/temperature gradient; this region is 
the CC, which increases in density and cools as it accretes more material and slows during its descent to the lower chromosphere.  The energy deposition within the CC is due to intermediate energy electrons in the beam.  Beam electrons at the low energy cutoff produce and heat the localized temperature increase in the chromosphere to $T=5-10$ MK. 

\item The layers below the CC can also be significantly heated by the high energy electrons in the beam ($E>>E_{\rm{c}}$ keV);  this region is referred to as the stationary flare layers because it exhibits negligible ($\lesssim 1$ km s$^{-1}$ upward) gas velocities relative to the CC.
The stationary flare layers extend several hundred km below the CC and are $T\sim9000-12,000$ K, which is less than the temperature range in the CC during its early evolution.  
\end{enumerate}

\begin{figure}[h!]
\plotone{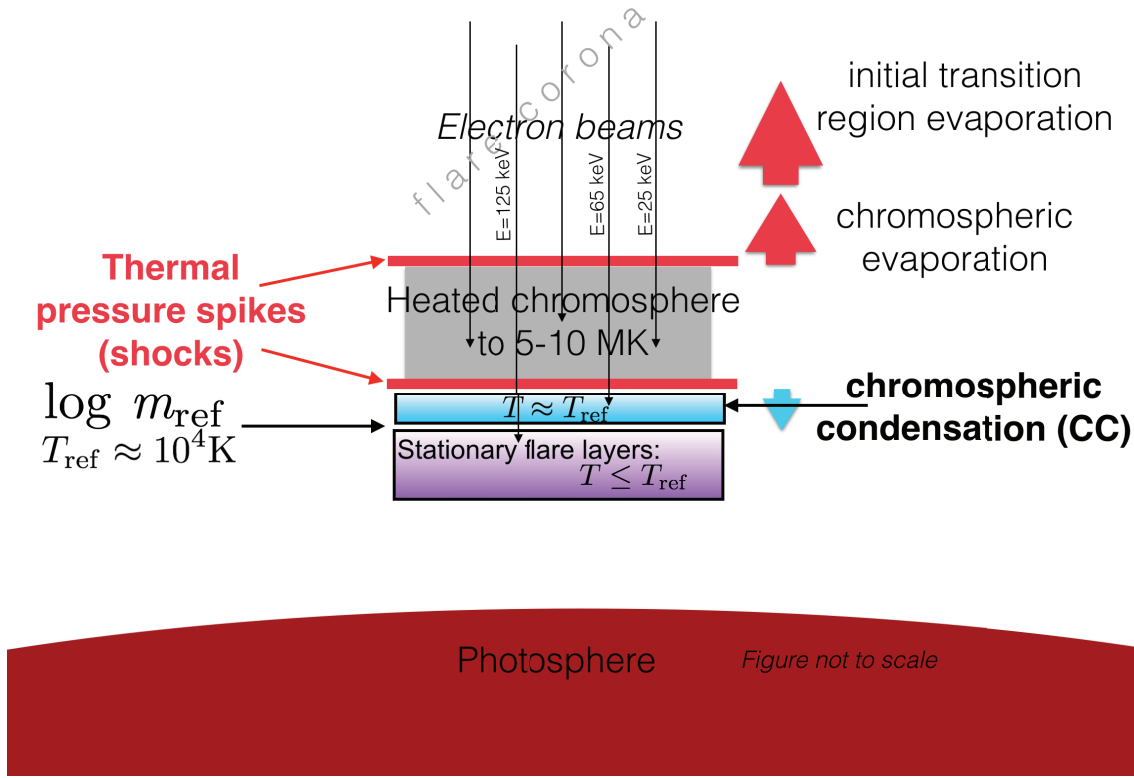}
%\plotone{atmos_evol_imp_0-2pt2s.eps}
\caption{Cartoon illustrating the important features that develop in the evolved atmospheric states with high beam flux heating rates.  The height corresponding to log $m_{\rm{ref}}$ and  $T_{\rm{ref}}$ occurs at the interface between the dense, downflowing cool ($T\approx10^4$ K) chromospheric condensation and heated stationary flare layers (note that by the evolved-1 time, the temperature at $m_{\rm{ref}}$ has increased to approximately $T_{\rm{ref}}+1500$ K for the very high beam flux density F13 models).  Thick colored arrows indicate plasma motions that have developed by the evolved times.  The approximate stopping depths for representative low (25 keV), intermediate (65 keV) and high energy (125 keV) beam electrons in the \solsoft\ model are indicated (black thin arrows) using the thick target formulae of \citet{Emslie1978, Emslie1981}, \citet{Ricchiazzi1983}, and \citet{HF94} corrected for relativistic length contraction \citep{Allred2015}.  For the \dmesoft\ model, representative electron beam energies for these arrows are 65 keV, 125 keV and 300 keV, respectively.  
Features in this figure have been adapted from \citet{Kowalski2016IAU}. }   \label{fig:cartoon}
\end{figure}

In Figure \ref{fig:cartoon}, we illustrate these two flaring layers that develop in response to heating by high flux density electron beams.

With extremely high beam energy flux densities ($\sim$F13), these two flaring regions develop an optical depth $\tau>1$ at
NUV and optical continuum wavelengths, and the emergent radiation is characterized by a hot $T\sim10^4$ K blackbody-like spectrum with a small Balmer jump ratio, as observed in spectral, broadband photometric, and narrowband photometric observations of dMe flares \citep{HP91, Zhilyaev2007, Fuhrmeister2008, Kowalski2013}.  If the beam energy flux densities are moderately high ($\sim$5F11), then bright NUV continuum intensity and Fe II emission lines are produced with a prominent red wing asymmetry, as observed in NUV (at $\lambda \sim 2830$ \AA) solar flare spectra from IRIS.  

Only after several seconds of high beam flux heating do these interesting properties develop in these simulations.  In this paper, we parameterize the temperature and density stratification of these RADYN simulations at the advanced states in order to make predictions and some general conclusions about the emergent continuum radiation spectrum over a large possible range of conditions in chromospheric condensations.  This will allow us (in future work) to select and run an interesting subset of RHD simulations based off of information at early times in order to make detailed line profile calculations at the evolved states of the simulations.

In the few high beam flux density simulations that have been completed with RADYN, we have noticed common patterns in their evolved states.  In this paper, these patterns are used for parameterizations of the evolved states of the temperature and density stratifications.    
The high beam flux density simulations that we use in this analysis are the 5F11 (extended heating run; ``\solsoft'') solar flare model from \citet{Kowalski2017A}, the F13 dMe flare model with a double-power (``\dmesoft'') distribution of electron energy \citep{Kowalski2015}, and the F13 dMe flare model with a harder single power law distribution (``\dmehard'') of electron energy with a power-law index of $\delta=3$ \citep{Kowalski2016}.  The model parameters are summarized in Table \ref{table:basics}.  The distinction between hard and soft beams is made according to the relative number of nonthermal electrons at $E \gtrsim 200$ keV, which are important for the heating and ionization in the stationary flare layers \citep{Kowalski2016}.  
  These RHD calculations 
were performed with the 1D RHD RADYN code \citep{Carlsson1992, Carlsson1994, Carlsson1995, Carlsson1997, Carlsson2002}, which calculates 
hydrogen, helium, and Ca II in NLTE and with non-equilibrium ionization/excitation (NEI).  We refer the reader to \citet{Allred2015}, \citet{Kowalski2015}, and \citet{Kowalski2017A} for extensive descriptions of the flare simulations. 

For each simulation, we analyze the atmospheric states at an ``early'' time and two ``evolved'' times in Table \ref{table:basics}.
The early times correspond to the early development of the CC when it exhibits a temperature of $T\sim40,000$ K and is flowing downward at approximately its maximum downflow speed;  this time also corresponds to when the stationary flare layers have achieved a temperature stratification that is relatively constant at any later time.  These conditions occur when the explosive temperature shock front in the chromosphere exceeds $T=2.5$ MK in the \solsoft\ and $T=10$ MK in the F13 models.  We choose $t=1.2$~s as the best time to represent the early state of the \solsoft\ model and $t=0.4$~s to represent the early states of the F13 models.    

The evolved times (``evolved-1'' and ``evolved-2'') in each simulation correspond to the times when the CC has cooled to $T\sim9000-13,000$ K and the stationary flare layers have heated to a similar temperate range.  At the evolved times in each simulation, the flare atmospheres produce the brightest optical and NUV continuum and line radiation as well as the largest continuum optical depth.  

Two evolved states in each RADYN simulation are considered to represent a range of possible extreme conditions that can be achieved.  For the \dmesoft\ model, the minimum temperature in the CC has decreased to $T\sim13,000$ K at $t=1.6$~s, which is also the time of the maximum emergent continuum intensity at wavelengths just shortward of the Balmer limit ($\lambda=3646$ \AA).  We refer to $t=1.6$~s as the evolved-1 time for the \dmesoft.  
However, the Balmer jump ratio continues to decrease to $t=2.2$~s as the CC accrues more material as it cools, resulting in a decrease of the physical depth range over which $\lambda=3500$ \AA\ photons escape \citep[see][]{Kowalski2015}.  At this point the blue $\lambda=4170$ \AA\ photons still escape from the stationary flare layers due to a lower optical depth in the CC $\tau_{4170}(\rm{CC})<1$.  The time of $t=2.2$~s is the evolved-2 time for the \dmesoft.  In the \dmehard\ simulation the CC cools to $T\sim13,000$ K at the evolved-1 time of $t=2.2$~s; the evolved-2 time is not attained before the heating ceases at $t=2.3$~s.  

For the \solsoft\ model,  the evolved-1 time is $t=3.97$~s, which was analyzed extensively in \citet{Kowalski2017A} and results in nearly the brightest NUV continuum intensity as the minimum temperature in the CC decreases to $T=10,000$ K.  The evolved-2 time corresponds to the maximum NUV continuum intensity at $t=5$~s.   

 The timescales for CC development in these high beam flux heating models are similar to the most recent observational constraints \citep[several seconds to twenty seconds;][]{Penn2016, Fatima2016} of electron beam heating duration in a single flare loop.  Therefore, the two evolved times bracket the possible range of atmospheric conditions, NUV continuum intensity, and NUV continuum optical depth in order to account for the uncertainty in the duration of flare heating in a loop. 
In summary, the evolved-1 times correspond to when the minimum temperature in the CC cools to $T\sim13,000$ K in high beam flux simulations and $T\sim10,000$ K in lower beam flux simulations.  The evolved-2 times correspond to further development of the CC at $\Delta t=0.6$~s after the evolved-1 time in very high (F13) beam flux simulations and at $\Delta t=1$~s after the evolved-1 time in lower (5F11) beam flux simulations.  

\floattable
\begin{deluxetable}{ccccccccc}
\rotate
\tabletypesize{\scriptsize}
\tablewidth{0pt}
\tablecaption{Model Parameters}
\tablehead{
\colhead{RADYN Model} & \colhead{Flux density} & \colhead{Heating duration (s)} & \colhead{log $g$} & \colhead{Power-law index ($\delta$)} & \colhead{Low energy cutoff (keV)} & \colhead{Early Time (s)} & \colhead{Evolved-1 Time (s)} & \colhead{Evolved-2 Time (s)} }
\startdata
 \solsoft\ & $5 \times 10^{11}$ & 15 & 4.44 & $4.2$ & 25 & 1.2 & 3.97 & 5.0 \\
 \dmesoft\ & $10^{13}$ & 2.3 & 4.75 & $3,4$ & 37 & 0.4 & 1.6 & 2.2 \\
 \dmehard\ & $10^{13}$ & 2.3  & 4.75 & $3$ & 37 & 0.4 & 2.2 &  -- \\
 \enddata
\tablecomments{  Basic information about the electron beam models and the times designated as the early, evolved-1, evolved-2 times for each.  The double power-law F13 model
has power-law indices of $\delta=3$ at $E<105$ keV and $\delta=4$ at $E>105$ keV.   The flux density above $E_{\rm{cutoff}}$ is given in units of erg cm$^{-2}$ s$^{-1}$.}
\label{table:basics}
\end{deluxetable}

\subsection{The Critical Flare Atmosphere Reference Parameters}

We use the \solsoft\ solar flare simulation to construct a simplified, approximate parameterization of the thermodynamic stratifications at the evolved-1 and evolved-2 times ($t=3.97$~s and 5~s, respectively) using only two reference atmospheric quantities in the RADYN calculation at the early time ($t=1.2$~s).  

\begin{figure}[h!]
\centering
\includegraphics[scale=0.44]{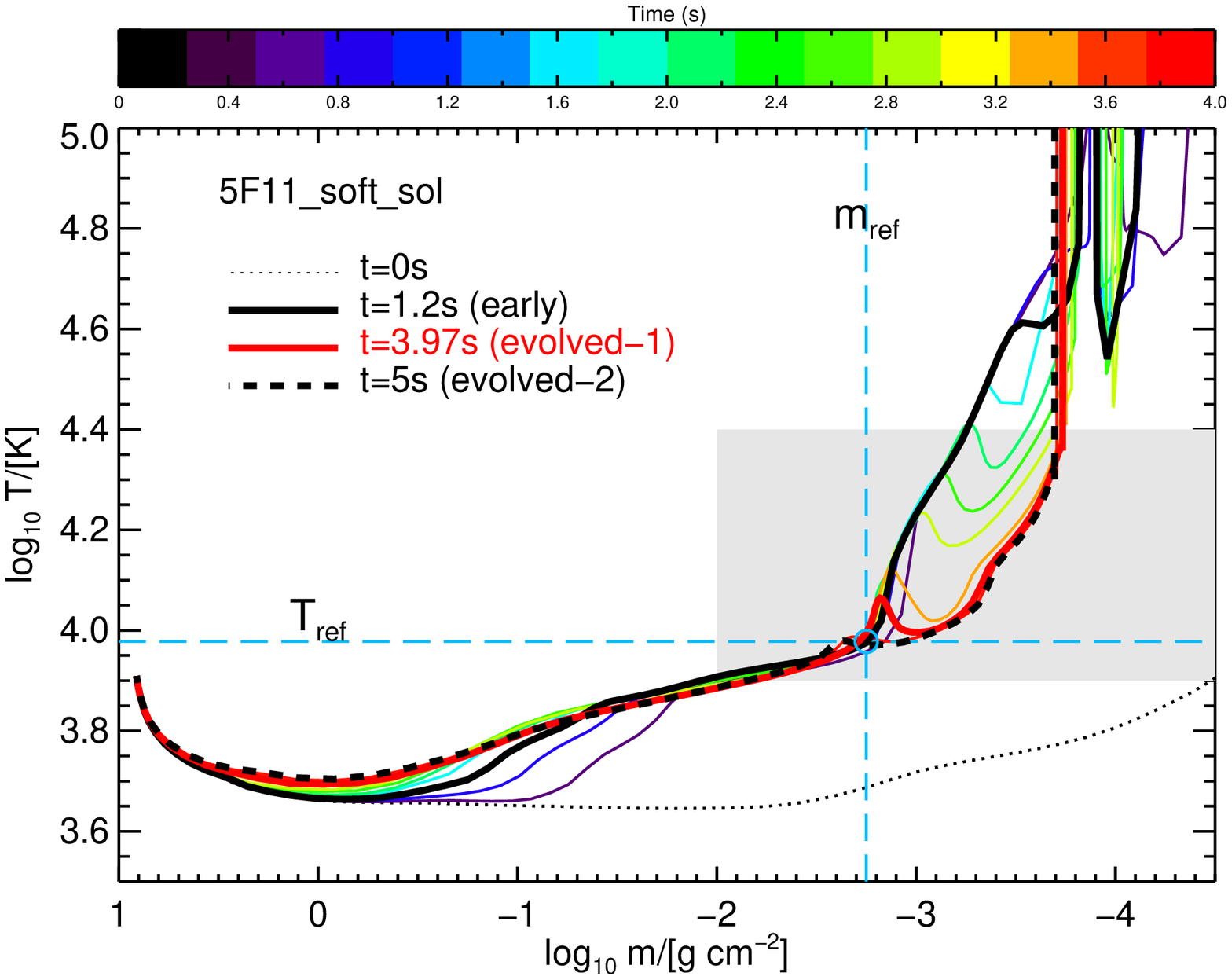}
\includegraphics[scale=0.44]{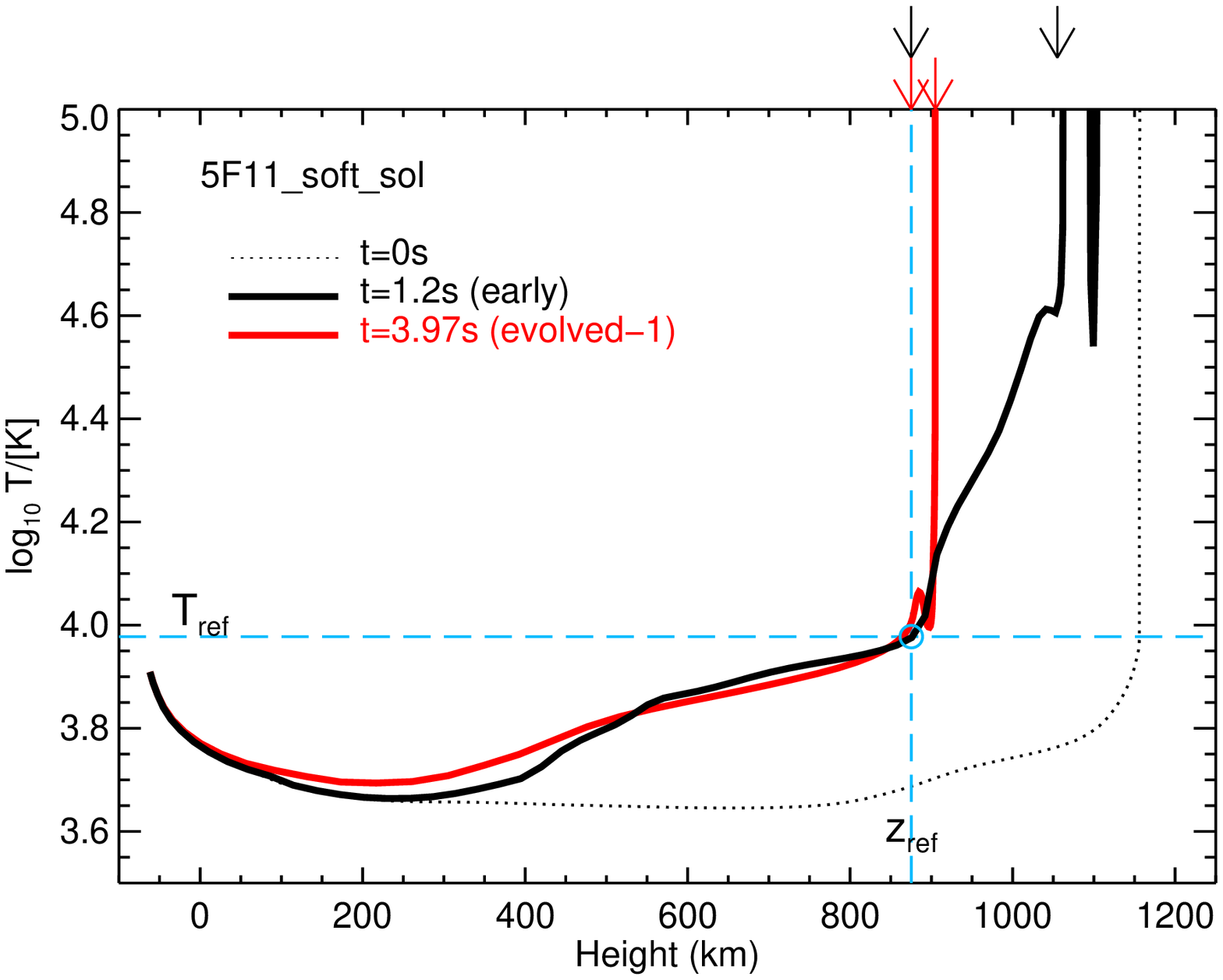}
\includegraphics[scale=0.44]{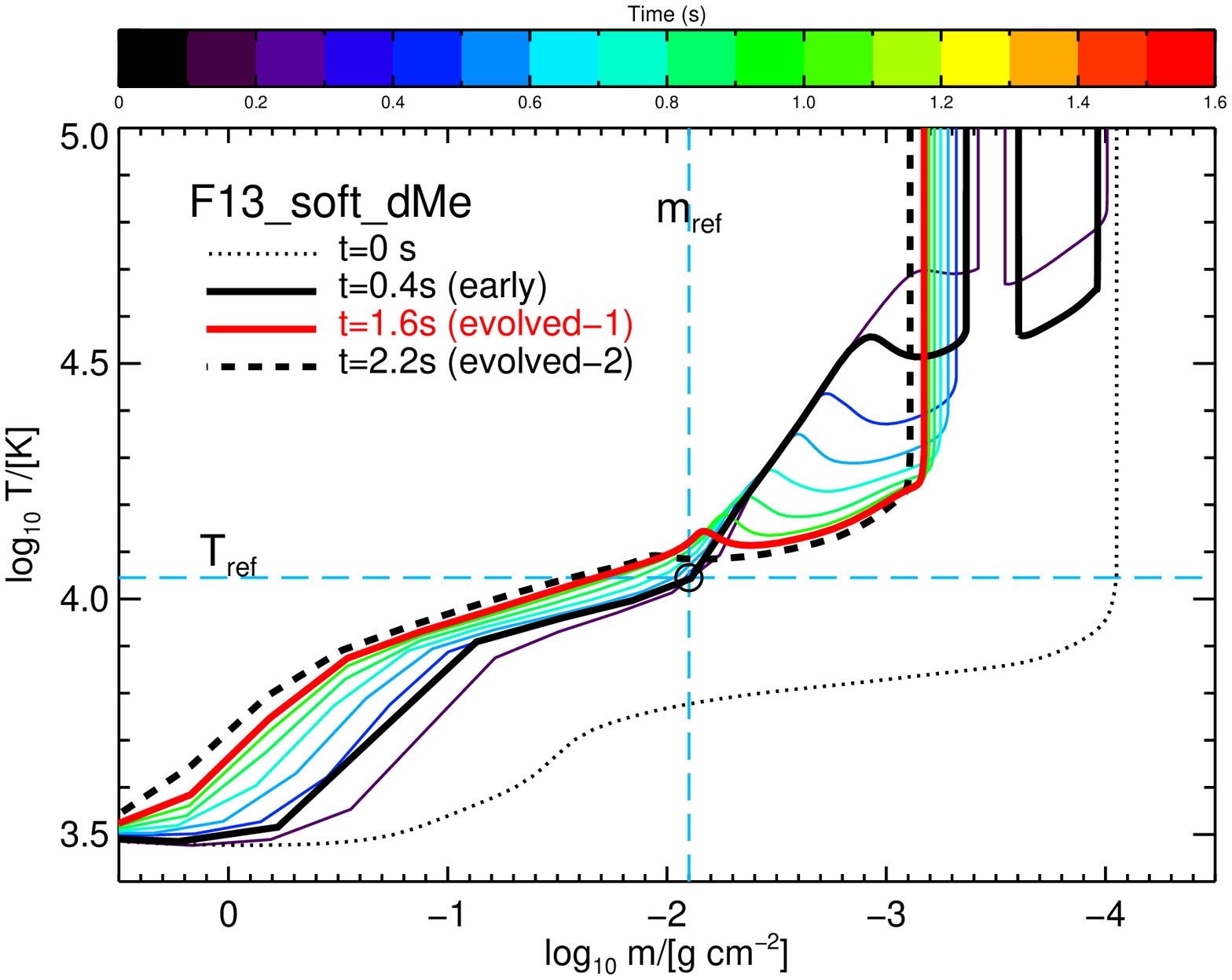}
\caption{Temperature evolution as a function of column mass (top) and height (middle) in the \solsoft\ model from $t=0-4$~s, shown at $\sim$0.4~s intervals and the evolved B time at $t=5$~s.  The vertical dashed line indicates the value of log $m_{\rm{ref}}=-2.75$, and the horizontal dashed line indicates the temperature $T_{\rm{ref}}=9500$ K.  A circle indicates the values of these parameters;  the middle panel clearly demonstrates that these values indicate the temperature gradient change at the early time from $T\lesssim 10,000$ K to $T\gtrsim 10,000$ K.  The arrows in the middle panel illustrate that the CC cools (from increased radiative losses) as it is compressed from a several hundred km region (between the black arrows) at the early time into a narrow 30 km region (between the red arrows) at the evolved-1 time.  (Bottom) The temperature evolution of the \dmesoft\ model for a dMe flare shown from $t=0-1.6$~s at 0.2~s intervals and at 2.2~s.  The evolved-1 atmosphere is indicated at $t=1.6$~s and the evolved-2 is indicated at $t=2.2$~s, when the maximum $\lambda=3500$ \AA\ continuum optical depth is achieved in the CC \citep{Kowalski2015}.  The gray shaded area in the top panel indicates the column mass and temperature ranges in Figure \ref{fig:5F11CC}. }   \label{fig:atmos_evol}
\end{figure}

Figure \ref{fig:atmos_evol} shows the temperature evolution of the \solsoft\ model from $t=0-3.97$~s.  After the early time of $t=$1.2~s, the atmospheric
temperature structure in the stationary flare layers at column mass $m$\footnote{We refer to column mass as log $m$ where $m$ has units of g cm$^{-2}$.} larger than log $m\gtrsim-2.75$ (corresponding to the vertical dashed blue line) does not change significantly (i.e., the thick black and thick red solid lines are similar at larger column mass than log $m\gtrsim-2.75$).
In the \solsoft\ simulation, the lower pressure gradient at the temperature explosion to $T=2.5-5$ MK compresses the gas that is initially spread over a physical depth range of $\Delta z \sim $180 km at the early time ($t=1.2$~s)
into a narrow region with a physical depth range of $\Delta z \sim 30$ km by $t=3.97$~s (the evolved-1 time).   This narrow 30 km region is the evolved CC.  

 The compression of gas into a CC can be seen in the middle panel of Figure \ref{fig:atmos_evol}, where we show the temperature stratifications at the early and evolved-1 times as a function of height.
 The ``flare transition region'' occurs at a steep pressure gradient where the temperature increases above the range shown on the y-axis in this figure; 
the flare transition region moves from $z\sim1075$ km at the early time to $z\sim 905$ km at the evolved-1 time.  This results in compression of the lower atmosphere between these two heights and thus an enhancement in the density in the CC by a factor of ten\footnote{The factor of ten enhancement in density is relative to the CC density at $t=1.2$~s and relative to the pre-flare density at the height of the CC at $t=3.97$~s.}.  The arrows at the top of the middle panel of Figure \ref{fig:atmos_evol} illustrate the physical depth ranges over which the atmosphere is compressed from the early to the evolved times.  The descent of the flare transition region to lower heights is a common feature of RHD simulations with hot coronae;  the flare transition region forms at a height where the density is such that the radiative losses balance the heat flux through the transition region. 

At the evolved-1 time, the bottom of the CC corresponds to the column mass of log $m=-2.75$, which occurs where the speed of the downflowing material falls below 5 km s$^{-1}$.  Furthermore, the temperature of the evolved-1 CC has decreased to a similar temperature as the top of the stationary flaring layers that are located just below the CC.   The properties of the CC at the evolved times when it is highly compressed and producing bright
continuum radiation can be predicted by determining the temperature and column mass at the top of the $T\sim10,000$ K stationary flare layers at an early time in the simulation\footnote{The stationary flare layers at the early time include material that is hotter than 10,000 K which gets accrued into the CC by the evolved times; it's interesting to note that the physical depth range of $\sim30$ km of the CC does not change much over the simulation.}. 
For the \solsoft\ model this temperature is $T=9,500$ K and this column mass is log $m=-2.75$. We denote these key reference parameters at early times as $T_{\rm{ref}}$ and  log $m_{\rm{ref}}$, respectively.  These values (at the early time, $t=1.2$~s) are indicated by light blue dashed lines in the top and middle panels of Figure \ref{fig:atmos_evol} for the \solsoft\ model.  The height of the critical reference parameters is indicated in the cartoon in Figure \ref{fig:cartoon}.

 In the bottom panel of Figure \ref{fig:atmos_evol} we show the temperature evolution for the \dmesoft\ simulation, which results in a value of log $m_{\rm{ref}} \sim -2.1$ that is deeper and a value of $T_{\rm{ref}} \sim 11,000$ K that is hotter than in the \solsoft\ solar flare model.  By the evolved-1 time ($t\sim1.6-1.7$~s) in the \dmesoft\ model, the CC has descended to the height and cooled to the temperature
of the top of the stationary flare layers, as in the \solsoft\ model.   Because log $m_{\rm{ref}}$ is larger than in the \solsoft\, more material has been accrued and compressed into the CC in the F13 model.  The values of $m_{\rm{ref}}$ and  $T_{\rm{ref}}$ for each model are given in Table \ref{table:ccparams}.

 Interestingly, there is a local temperature maximum in all RADYN
models (Figure \ref{fig:atmos_evol}) that occurs just to lower column mass than $m_{\rm{ref}}$ at the
evolved-1 time.  This relatively small temperature increase is also located just to higher column mass than
$m_{\rm{ref}}$ at the evolved-2 time.  Thus, the evolved-2 and evolved-1 times can be
consistently identified in any simulation if the local temperature
maximum straddles $m_{\rm{ref}}$ at these two times.  

The values of $T_{\rm{ref}}$ and  log $m_{\rm{ref}}$ denote a meaningful change in the temperature gradient at the early times:
the value of log $m_{\rm{ref}}$ demarcates the height ($z_{\rm{ref}}$; Figure \ref{fig:atmos_evol} middle panel) below which temperature is nearly constant at $T \lesssim 10^4$ K and above which the temperature rises steeply to the temperature of the $T\sim40,000$ K CC, before rising again to $T>100,000$ K in the narrow flare transition region.
The location of $m_{\rm{ref}}$ occurs at the height where the hydrogen ionization
fraction increases from $X_{\rm{ion}}=80-90$\% to $X_{\rm{ion}}=99.9$\%, which results in the large
gradient in temperature up to a plateau with $T=40,000 - 60,000$ K at the early times. 

A simplified analysis of the energy balance at the early times in a RADYN simulation reveals the physical origin of $m_{\rm{ref}}$ and identifies its approximate value.  We define the approximate capacity of hydrogen in an atmosphere to regulate beam heating as

\begin{equation} \label{eq:radiative}
ie_{\rm{hydrogen}}(z,t=0) \approx 13.58 \rm{eV}\ \times n_{\rm{HI},n=1}(z,t=0) 
\end{equation}

\noindent which is the total ionization energy ($ie$) of hydrogen at atmospheric height $z$ in the pre-flare atmosphere.  In the pre-flare atmosphere, Equation \ref{eq:radiative} sensibly decreases towards increasing heights as the density of
hydrogen drops.  The integral $\int^{\rm{t_{early}}}_0 Q_{\rm{beam}}(z,t) dt$ is the cumulative energy deposited by the nonthermal electron beam from $t=0$~s to the early time.  $Q_{\rm{beam}}(z,t)$ (the beam energy deposition rate) decreases towards lower heights, and
 the intersection of the two curves $\int^{\rm{t_{early}}}_0 Q_{\rm{beam}}(z) dt$ and $ie_{\rm{hydrogen}}(z,t=0)$ indicates the
approximate value of $m_{\rm{ref}}$ for all three models in Table \ref{table:basics}.  Thus, $m_{\rm{ref}}$ indicates where hydrogen transitions from partial ionization below $m_{\rm{ref}}$ to nearly complete ionization at the heights above $m_{\rm{ref}}$.
The atmosphere heats in response to the beam energy, and there is additional cooling from (primarily) hydrogen Balmer and Paschen transitions at the depths where these curves intersect.  Thus, to $ie$ one can add the net time- and wavelength-integrated cooling from $t=0$~s to the early time for hydrogen transitions to obtain a closer estimate of  $m_{\rm{ref}}$.

  As expected, a factor of 20 higher beam flux density in the F13 models results in more column mass of hydrogen being fully (99.9\%) ionized than in the \solsoft\ model and thus larger values of $m_{\rm{ref}}$. Between the two F13 models, the \dmehard\  has the harder electron beam distribution \citep[with more nonthermal electron energy at $E>200$ keV; see discussion in][]{Kowalski2016}, a slightly larger value of $m_{\rm{ref}}$, and a slightly higher value of $T_{\rm{ref}}$ than the \dmesoft\ (Table \ref{table:ccparams}).  The energy flux density in the high-energy electrons ($E\gtrsim200$ keV) in these beams and thus the beam hardness and total flux density determine how deep hydrogen is completely ionized and can no longer regulate heating from electron beam energy deposition.

\subsection{Predicting the CC Evolution from $T_{\rm{ref}}$ and $m_{\rm{ref}}$} \label{sec:method_param}
Larger values of $m_{\rm{ref}}$ produce larger continuum optical depth and larger emergent continuum intensity in the flare atmosphere.  
The maximum density in the evolved CC in the \dmesoft\ RADYN model is $n_{\rm{H}}=7\times10^{15}$ cm$^{-3}$ whereas the maximum density in the evolved CC in the \solsoft\ RADYN model is $5\times10^{14}$ cm$^{-3}$.  As a result, the NUV (at $\lambda=3500$ \AA) continuum optical depth in the CC in the F13 model is large, $\tau_{3500\rm{AA}}(\rm{CC}) \sim5$, whereas the NUV continuum optical depth in the CC in the \solsoft\ model is smaller, $\tau_{3500\rm{AA}}(\rm{CC})=0.1$.   
Scaling relationships from $m_{\rm{ref}}$ (given that $T_{\rm{ref}}$ always occurs at $T\approx 10,000$ K) would be invaluable for comparison to NUV and blue spectral observations of flares in order to constrain the optical depth and electron density (Section \ref{sec:application}).

\begin{figure}[h!]
\plotone{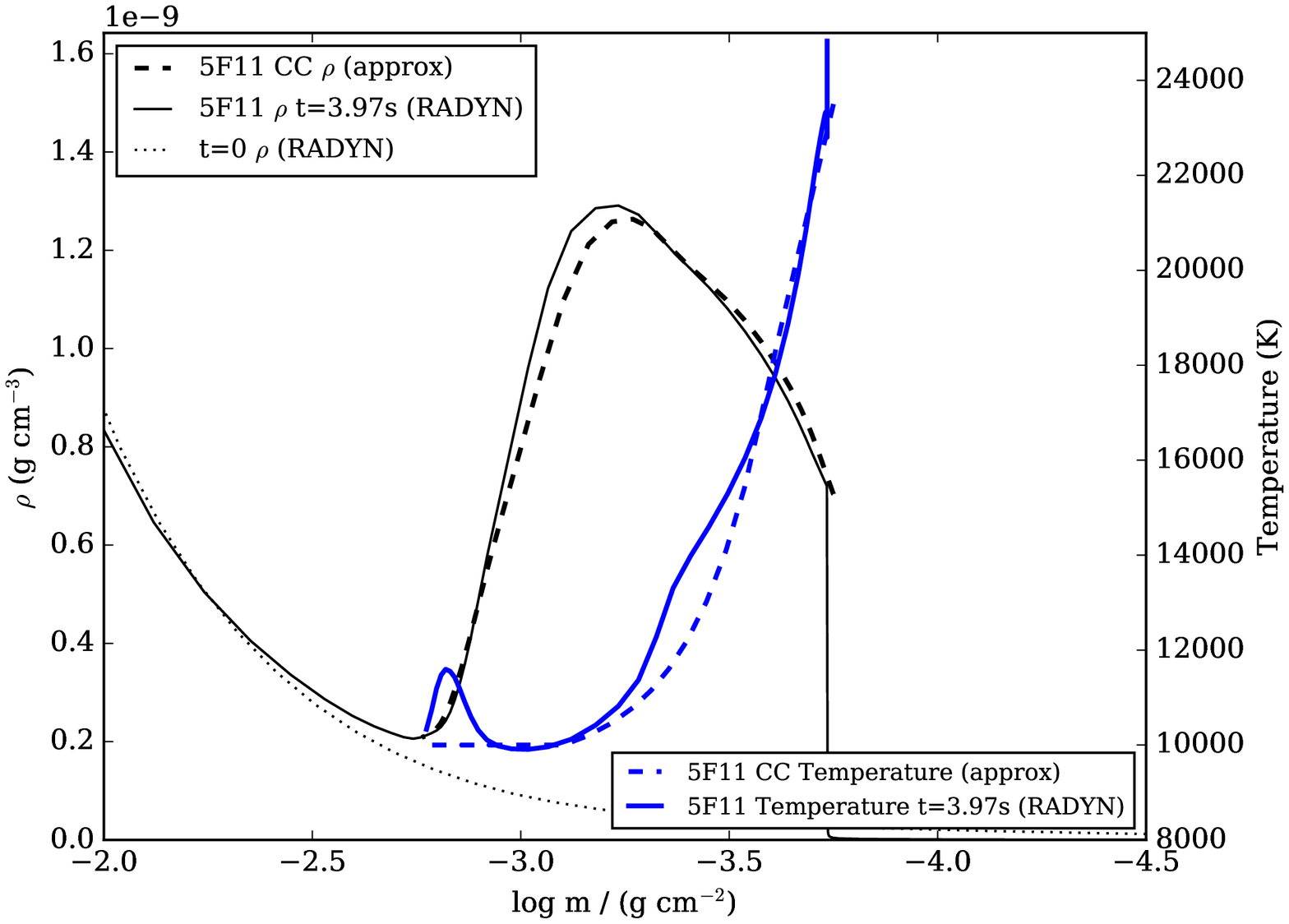}
\caption{ Enlarged view of the chromospheric condensation (CC) at the evolved-1 time ($t=3.97$~s) in the \solsoft\ model at column mass corresponding to the gray shaded region in Figure \ref{fig:atmos_evol}.  The density stratification at $t=0$~s in 
the solar atmosphere is shown as a dotted line for comparison.  
The approximate model of the CC density and temperature stratification at the evolved-1 time are shown as dashed lines.  The parameters used for this are $x_{\rm{maxCC}}=30$ km, log $m_{\rm{ref}}=-2.75$, and $T_{\rm{ref}}=9500$ K, and $T_{\rm{minCC}}=10,000$ K (see text).}   \label{fig:5F11CC}
\end{figure}

Using the RADYN calculation of the atmospheric response to a 5F11 beam flux density, we present a method to estimate the white-light continuum optical depth in evolved CCs and the emergent continuum intensity from the two flaring layers at
the evolved times using only the parameters $T_{\rm{ref}}$ and log $m_{\rm{ref}}$ at early times.  
 In this section we present the parameterization of the CC and the top of the stationary flare layers;  in Section \ref{sec:stationary} we present the parameterized stratification of the stationary flare layers.   Additional details are given in Appendix A.

To construct approximate evolved states of the CC, we take the following steps:

\begin{enumerate}

 \item We obtain values of log $m_{\rm{ref}}$ and $T_{\rm{ref}}$ at early times in the RADYN simulations of Table \ref{table:basics}.  For very high (F13) beam flux density simulations, a time of $t=0.4$~s is adequate, whereas for lower beam flux density simulations such as for the \solsoft\, $t=1.2$~s is adequate.  
 To obtain log $m_{\rm{ref}}$ and $T_{\rm{ref}}$ consistently for any set of models at their early times, we calculate the quantity $d$ log $T$/$d$ log $m$ and find the column mass where this quantity increases above $-0.3$ at the early times.  At larger column mass than $m_{\rm{ref}}$,
this derivative is near 0.  At lower column mass, this derivative has much more negative values of $<-0.3$.
 The values of log $m_{\rm{ref}}$ and $T_{\rm{ref}}$ obtained in this way for each model are given in Table \ref{table:ccparams}.  

\item At the evolved-1 time of the \solsoft\, we  
 obtain the temperature, velocity, mass density, and ionization fraction 
stratification for the regions of the atmosphere corresponding to temperatures $T\lesssim25,000$ K and where $v_z < -5$ km s$^{-1}$.  This region of the atmosphere corresponds to the cool, dense region of the CC\footnote{Higher temperatures at higher heights and high speeds $\sim50$ km s$^{-1}$ are also downflowing and are thus part of the CC at the early and evolved times.   We do not consider temperatures higher than 25,000 K at the evolved times because the lower densities at these temperatures do not appreciably contribute to the emergent
NUV and optical continuum radiation. }.  We define $z_o$ as the height corresponding to $T\approx 25,000$ K, where $z$ is the height variable from RADYN and $z=0$ occurs at $\tau_{5000\rm{AA}}=1$.  
We define the distance from the top of the CC to any lower height $z$ as $x=|z-z_o|$, where $x$ increases toward the pre-flare photosphere.  For the \solsoft\ model, we set $x_{\rm{maxCC}} = 30$ km as the maximum physical depth range of the evolved CC as in the RADYN simulation. 

The density stratification of the evolved CCs in the \dmehard\ and \dmesoft\ models are qualitatively similar to the evolved CC in the \solsoft\ model, but they have a larger value of the maximum mass density ($\rho_{\rm{maxCC}}$) and a smaller 
physical depth range of $x_{\rm{maxCC}} \approx 18$ km.  Compared to 30 km, a depth range of 18 km is very close to what the ratio of the surface gravities indicates for the physical depth range ($x_{\rm{maxCC}}$) of a compression in a dMe atmosphere with a higher gravity of log $g=4.75$.  The CC in the solar atmosphere is more extended over height and exhibits a lower $\rho_{\rm{maxCC}}$ because of the lower surface gravity by a factor of two.   In the solar CC, the lower mass density also results from a smaller amount of material that is compressed in the CC due to a smaller value of log $m_{\rm{ref}}$.  
The maximum density attained in a CC can also be affected by the velocity field such that much larger velocity gradients than in the \solsoft\ model may produce a different density stratification in the CC; we discuss the role of this parameter for a higher electron beam flux density solar flare model in Section \ref{sec:superflares}.

Using the density stratification of the CC in the \solsoft\ model at the evolved-1 time as a template, we create an approximate density stratification for any CC at the evolved-1 and evolved-2 times by applying values of log $m_{\rm{ref}}$, $T_{\rm{ref}}$, and log $g$ obtained at the early time. The advantage of this is that it predicts the evolved states directly from the early state without the expensive computations required to actually evolve the RADYN simulations.  The CC density template ($\rho(x)_{\rm{norm}}$) is obtained by normalizing the density stratification of the CC at the evolved-1 time in the \solsoft\ model by its maximum density ($\rho_{\rm{maxCC}}=1.3\times10^{-9}$ g cm$^{-3}$). This density stratification is plotted in Figure \ref{fig:5F11CC}.

The CC density stratification template extends from the location where the speed of downflowing material falls below 5 km s$^{-1}$ (at the low temperature end, lower height end of the CC) to the location where the temperature exceeds $T=25,000$ K (at greater heights in the CC).  The column mass at the low temperature end of the normalized density stratification ($\rho(x)_{\rm{norm}}$) is set to the value of $m_{\rm{ref}}$, and the height scale ($dx_{\rm{Sol}}$)
of the template is adjusted according to the surface gravity.  We solve the equation

\begin{equation} \label{eq:template}
C \int_{x=0}^{x=30\rm{km}} \rho(x)_{\rm{norm}} \frac{10^{4.44}}{10^{\rm{log} g}} dx_{\rm{Sol}}=m_{\rm{ref}}
\end{equation}

\noindent for the constant $C$ to obtain a density stratification $C \rho(x)_{\rm{norm}} $ with units of g cm$^{-3}$ on a height scale 
$\frac{10^{4.44}}{10^{\rm{log} g}} dx_{\rm{Sol}}$ in units of cm, where $x_{\rm{maxCC}}=\int_{x=0}^{30\rm{km}}\frac{10^{4.44}}{10^{\rm{log} g}} dx_{\rm{Sol}}$.  Before solving for $C$, we subtract 10\% from $m_{\rm{ref}}$ in order to account for mass evaporated into the corona\footnote{This value of $\sim10$\% is obtained in the RADYN simulations and does not correspond to the fraction of beam energy that goes into evaporation;  for the \solsoft\ model, $\sim50$\% of the beam energy is deposited higher than the CC because cutoff energy electrons are stopped higher (see Figure \ref{fig:cartoon} here and Table 3 of \citet{Kowalski2017A}).}.  The approximate evolved-1 atmosphere for the 5F11 compared to the RADYN calculation is shown in Figure \ref{fig:5F11CC}.  The approximate evolved-1 atmosphere for the \dmehard\ using the values of $m_{\rm{ref}}$ and $T_{\rm{ref}}$ in Table \ref{table:ccparams} and log $g=4.75$ ($x_{\rm{maxCC}}=15$ km) is shown in Figure \ref{fig:F13CC} compared to the RADYN calculation.  There is satisfactory agreement in the peak density and the general shape of the density stratification.  At $x>2$ km, there is an exponential decay of the density from $\rho_{\rm{maxCC}}$ to the stationary flare layers in the RADYN calculation, and the approximate evolved-1 stratification exhibits a steeper decrease (smaller scale height) than in the RADYN calculation.  This discrepancy is discussed further in Section \ref{sec:superflares}.

\begin{figure}[h!]
\plotone{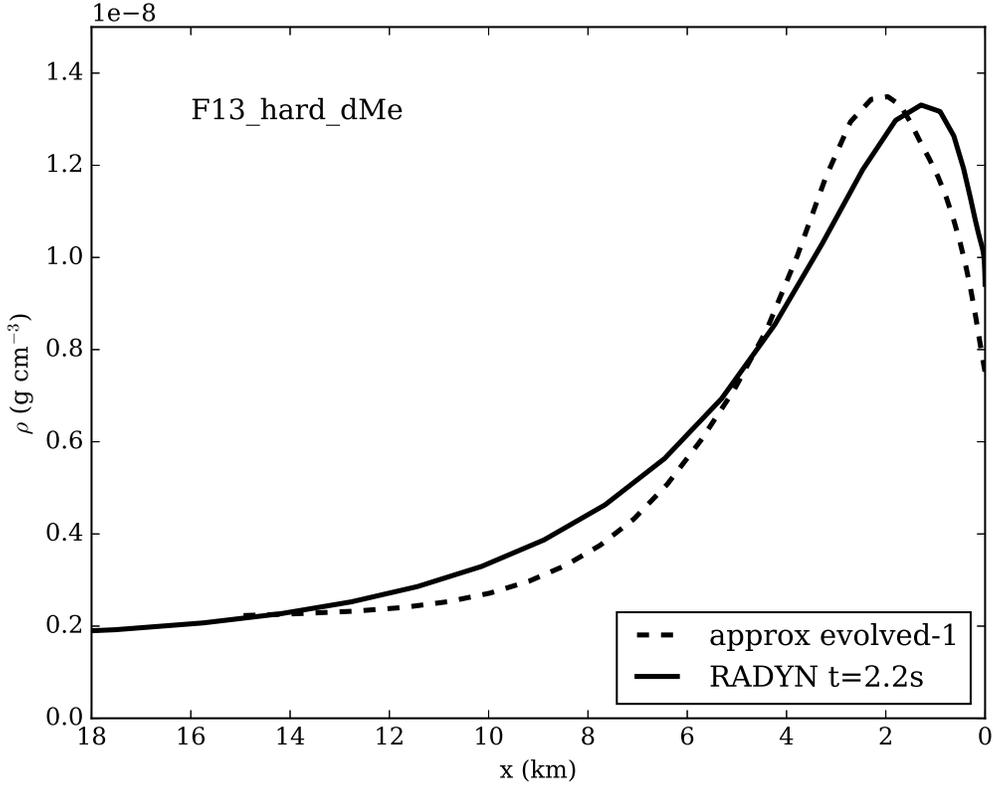}
\caption{ The density stratification in the approximation at the evolved-1 time of the \dmehard\ model compared to the RADYN calculation at $t=2.2$~s.  The top of the low-temperature ($T<25,000$ K) region of the CC is at $x=0$ km, and the bottom of the CC is at $x=x_{\rm{maxCC}}=15$ km.  The density stratification was calculated from the density stratification template in Figure \ref{fig:5F11CC} using the value of log $m_{\rm{ref}}=-2.04$ in Equation \ref{eq:template}.}   \label{fig:F13CC}
\end{figure}

\item Our template CC requires a temperature stratification, which we also obtain from the \solsoft\ model.
The minimum temperature ($T_{\rm{min CC}}$) in a CC at the evolved-1 time occurs at the height where $\rho \sim 3/4 \times \rho_{\rm{maxCC}}$ at the higher column mass end of the CC. In Figure \ref{fig:5F11CC}, $T_{\rm{min CC}}=10,000$ K occurs at log $m \sim -3.0$ and is near the value of $T_{\rm{ref}}=9500$ K: the CC has cooled to a temperature that is similar to the temperature at the top of the stationary flare layers.  The value of $m_{\rm{ref}}$ at the evolved-1 time shifts to lower column mass (log$m\sim -3.3$ in the \solsoft\ model) because most of the CC has cooled to $T\sim 10,000$ K and a significant fraction of hydrogen is not ionized in the evolved CC.  The temperature stratification vs. column mass is qualitatively similar in the CCs among the \solsoft\ and F13 models, but the  
the value of $T_{\rm{minCC}}$ is higher in the F13 models.  

We use the \solsoft, \dmesoft, and \dmehard\ models to prescribe simple adjustments to the temperature at the top of the stationary flare layers and the minimum temperature of the CC ($T_{\rm{min CC}}$)) because these temperatures are approximately equal at the evolved times.  
For the high beam flux density simulations, the temperatures in the stationary flare layers at $m>m_{\rm{ref}}$ increase by $\Delta T \sim 1500-2000$ K from the early to the evolved-1 times, and for the \solsoft\ model the temperature at the bottom of the CC at the evolved-1 time is $\Delta T \sim 500$ K higher than the temperature indicated by $T_{\rm{ref}}$ at the early time.  
The amount by which the stationary flare layers increase in temperature through the simulation is sensitive to the hardness and flux of the electron beam distribution:  harder beams and higher fluxes result in more heating and a higher (thermal) ionization fraction of hydrogen of the stationary flare layers \citep{Kowalski2016}.   In the approximate evolved-1 model atmospheres, we simply take 
either $T_{\rm{minCC}}$(evolved-1)$=T_{\rm{ref}}\rm{(early)}+500$ K for the lower beam flux density (5F11) models or $T_{\rm{minCC}}$(evolved-1)$=T_{\rm{ref}}\rm{(early)}+1500$ K for the higher beam flux density (F13) models.  The temperature at the top of the stationary flare layers at the evolved-1 times is set to $T_{\rm{minCC}}$.  

At the evolved-2 times in the RADYN simulations, the bottom of the CC at $x_{\rm{maxCC}}$ descends to $\sim1.5$ times higher column mass than $m_{\rm{ref}}$.  The maximum emergent continuum intensity occurs in the \solsoft\ model, and the maximum continuum optical depth occurs in the \dmesoft\ model.  At the evolved-2 times, the values of $T_{\rm{minCC}}$ are $\sim500$ K less than the values at the evolved-1 times because the CC has increased in density further and thus experiences more radiative cooling.    
 We find that at the evolved-2 times in the RADYN simulations, the value of $m_{\rm{ref}}$ occurs where the density stratification of the CC decreases to $0.45$ times the maximum density in the CC.  The approximate density stratification models at the evolved-2 times are calculated by evaluating Equation \ref{eq:template} with the upper limit of integration set to $x=10$ km, which is where $\rho(x)=0.45\rho_{\rm{maxCC}}$.  At the evolved-2 times, the value of $x_{\rm{maxCC}}$ does not change, but the value of $m_{\rm{ref}}$ occurs at  $\frac{1}{3}x_{\rm{maxCC}}$.  We set
 $T_{\rm{minCC}}$(evolved-2)$=T_{\rm{ref}}\rm{(early)}$ for lower beam flux density (5F11) models and $T_{\rm{minCC}}$(evolved-2)$=T_{\rm{ref}}\rm{(early)}+1000$ K for higher beam flux density (F13) models.  The temperature at the top of the stationary flare layers at the evolved-2 times is set to $T_{\rm{minCC}}$, as for the evolved-1 times.  

The details for establishing the temperature stratification at higher and lower heights than the height corresponding to 
$T_{\rm{minCC}}$ are presented in Appendix A.  
The approximate evolved-1 temperature stratification is shown in Figure \ref{fig:5F11CC} compared to the \solsoft\ RADYN calculation.

  \item To calculate the continuum optical depth within the approximate, evolved CC, we use LTE population densities of hydrogen and the H-minus ion. The evolved CCs become very dense in the RADYN simulations, and the hydrogen level populations are close to LTE values except at the upper $\sim1$ km of the CC where the $n=1$ and $n=2$ populations depart significantly from their equilibrium values.  From the mass density stratification of the CC, we convert to $n_{\rm{H, tot}}(x)$ using the gram per hydrogen value of 2.269$\times10^{-24}$ for the solar abundance.  From the temperature stratification of our approximate model CCs, we use the Saha-Boltzmann equation to solve for the hydrogen ionization fraction and the level populations as a function of height.   We solve for the LTE electron density first by truncating the hydrogen atoms with $n_{\rm{max}}=10$.  This approximate electron density is used to solve for the partition function and level population densities of hydrogen using the occupational probability formalism of \citet{HM88} with $n_{\rm{max}}=100$.  Then, the we re-solve for the LTE electron density.

  \item The equations for the hydrogen bound-free opacity, hydrogen free-free opacity, and H-minus bound-free opacity are used to calculate the continuum optical depth, at the base of the approximate, evolved CC, $\tau_{\lambda}(\rm{CC})$,  using $x_{\rm{maxCC}}=15$ km for the dMe atmosphere and $x_{\rm{maxCC}}=30$ km for the solar atmosphere.   Continuum opacities are corrected for stimulated emission.

The results for several NUV and optical continuum wavelengths ($\lambda=2826, 3500, 4170, $ and 6010 \AA) are shown in Table \ref{table:ccparams} compared to the optical depth calculated using the NLTE, NEI populations from RADYN and the continuum optical depth calculation method in \citet{Kowalski2017A}. The NUV continuum wavelength $\lambda=2826$ \AA\ corresponds to the NUV wavelengths observed by the Interface Region Imaging Spectrograph \citep[IRIS;][and see section \ref{sec:iris} here]{DePontieu2014}.  The continuum wavelengths $\lambda=3500, 4170, 6010$ \AA\ correspond to the central wavelengths of custom filters in the NUV, blue, and red, respectively, used to study dMe flares with the 
ULTRACAM instrument \citep[][and see section \ref{sec:relation} here]{Dhillon2007, Kowalski2016}.
The evolved-1 and the evolved-2 approximations are in excellent agreement with the 
continuum optical depth values at the bottom of the CCs at the respective times in the RADYN calculations.   In the last two columns, we show that the maximum LTE electron density values in the approximate model CCs are generally consistent with the NLTE, NEI calculation from RADYN.  We also show our predictions for the \dmehard\ model at the evolved-2 time, which is not calculated in RADYN simulation.  

\end{enumerate}

\subsubsection{Approximating the Heating in the Stationary Flare Layers} \label{sec:stationary}
To calculate the approximate emergent specific continuum intensity and emergent specific radiative flux density for comparison to observed flare spectra, we construct a simplified representation of the layers below the CC that are heated by the beam electrons with $E>>E_{\rm{c}}$\footnote{For the F13 models, $E \gtrsim 200$ keV electrons heat the column mass greater than log $m_{\rm{ref}}$ and for the \solsoft\ model, $E \gtrsim 80$ keV electrons heat the layers at column mass greater than log $m_{\rm{ref}}$ at the evolved times;  see Figure \ref{fig:cartoon}.}.  These stationary flare layers can contribute significantly to the emergent continuum radiation at $\lambda$ if the optical depth in the CC is $\tau_{\lambda}(\rm{CC})<1$;  if $\tau_{\lambda}(\rm{CC})>1$ at some continuum wavelengths,
the spectral shape of the emergent intensity will be modified from the spectral energy distribution that is expected from hydrogen recombination emissivity \citep{Kowalski2015}. 

For the density stratification of the stationary flare layers, we either choose the solar or dMe pre-flare density stratification since material does not compress at these heights.  We join the density stratification of the stationary flare layers to the CC to form a continuous density stratification in our approximate, evolved flare atmosphere.   
The details of the temperature stratification for the stationary flare layers is presented in Appendix A.2.  In summary, the temperature
decreases from  $T_{\rm{minCC}}$ (at the top of the stationary flare layers) to  $T_{\rm{minCC}}-3000$ K (at the bottom of the stationary flare layers) for 
hard ($\delta \sim 3$) beam models and to  $T_{\rm{minCC}}-5000$ K (at the bottom of the stationary flare layers) for 
soft ($\delta \gtrsim 4$) beam models.  The electron density in the stationary flare layers is determined
under the LTE conditions from the given temperature stratification.  We calculate the LTE populations of hydrogen and the H-minus ion and the continuum emissivity in the stationary flare layers as done in the CC (Section \ref{sec:method_param}).  

\subsubsection{Emergent Continuum Spectra}
\begin{figure}[h!]
\plotone{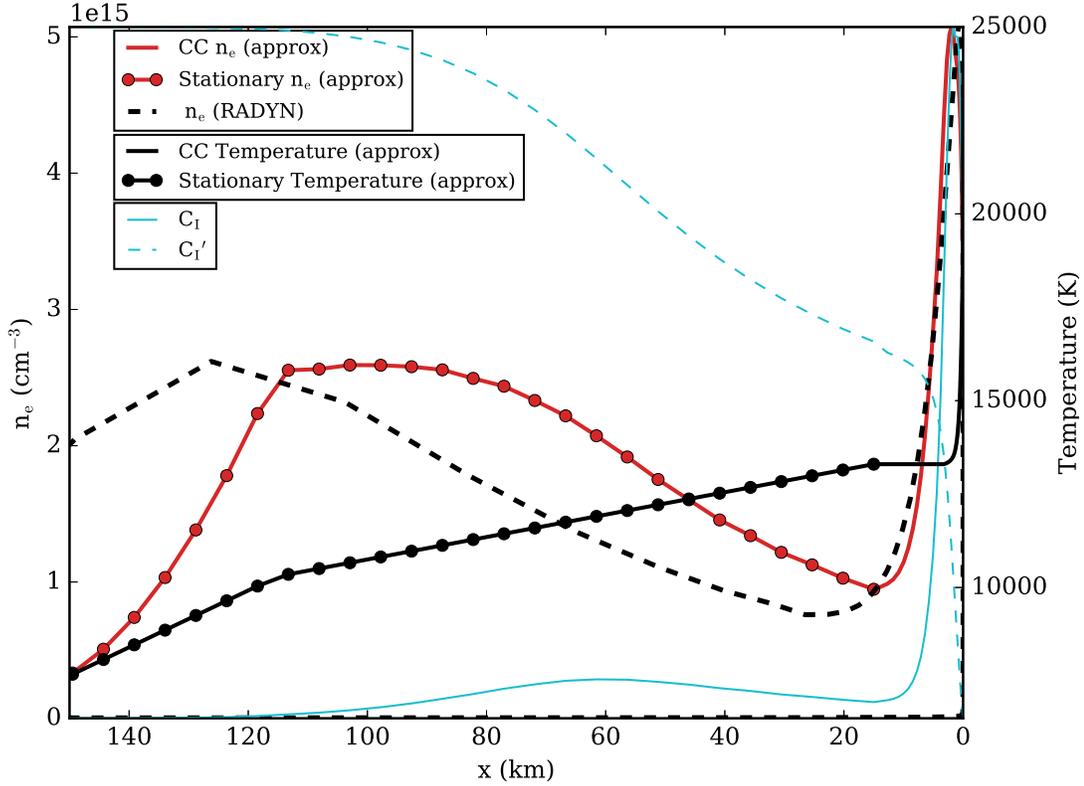}
\caption{The RADYN calculation of the electron density at the evolved-1 time of the \dmehard\ model compared to our approximations.  The temperature stratification from our approximations are also shown.  The top of the CC corresponds to $x=0$ km and extends to $x=15$ km.   The contribution function to the emergent blue continuum (4170 \AA) intensity is shown as the solid light blue line normalized to the peak value, and the cumulative contribution function ($C_I^{\prime}$) is shown as the dashed light blue raning from 0 to 1 on the right axis.  The approximate model for the evolved-1 time predicts the correct fraction (0.45) of the emergent blue continuum intensity originating from the stationary flare layers.  In the RADYN simulation, the density decreases more than in the approximate model before the onset of the stationary flare layers, which results in the offset in the electron density to lower heights (larger values of $x$).  However, the overall shape and magnitude of the electron density stratification is well reproduced with our approximations.  }   \label{fig:F13_evolvedA}
\end{figure}

The approximate evolved-1 temperature and electron density stratification for the \dmehard\  model is shown in Figure \ref{fig:F13_evolvedA} compared to the RADYN calculation of the electron density.   Within the CC, the gas density is well-reproduced (see Figure \ref{fig:F13CC}).  In the stationary flare layers, the maximum electron density and the electron density stratification is well-reproduced but the location of the maximum is offset towards greater heights.  This discrepancy is a result of our simple way of appending the density stratification of the stationary flare layers to the evolved CC model.   

 Multiplying the emissivity at all heights by $e^{-\tau_{\lambda}/\mu}/\mu$ and integrating over height gives the emergent continuum intensity from the simplified, evolved model atmospheres.  We calculate the cumulative contribution function \citep[$C_I^{\prime}$;][]{Kowalski2017A} which allows us determine the fraction of emergent intensity originating from a height greater than $z$.  The cumulative contribution function at $\lambda=4170$ \AA\ is shown for our approximate model of the \dmehard\ simulation at the evolved-1 time in Figure \ref{fig:F13_evolvedA}.  The physical depth range from $C_I^{\prime}=0.05$ to 0.95 for the emergent blue continuum intensity is $\Delta z =86$ km (vs. 95 km in RADYN), the 
fraction of emergent blue continuum intensity from the stationary flare layers is 0.46 (vs. 0.45 in RADYN), and the FWHM of the contribution function in the CC is 3.7 km (vs. 2.2 km in RADYN).  Our approximate evolved-1 model atmosphere also satisfactorily reproduces the moderate ($\tau_{4170}(\rm{CC}) \sim 0.5$) blue continuum optical depth in the CC (vs. $\tau_{4170}(\rm{CC}) \sim 0.6$; Table \ref{table:ccparams}), which is critical for producing the observed $T\sim10,000$ K blackbody-like continua in the emergent radiative flux in these models \citep{Kowalski2015, Kowalski2016}.  

We calculate the emergent specific radiative flux density, $F_{\lambda}$, using a Gaussian integral with the same five outgoing $\mu$ values employed in RADYN, in order to compare to unresolved stellar observations.  
The results for the Balmer jump ratio, $F_{\lambda=3500}/F_{\lambda=4170}$ (FcolorB), in the emergent radiative flux spectra compared to the RADYN calculations in \citet{Kowalski2016} are shown in Table \ref{table:ccparams}.  Large Balmer jump ratios of FcolorB$>8$ are produced in the \solsoft\ model and the evolved approximations, whereas small Balmer jump ratios of FcolorB$\lesssim2$
are produced in the F13 RADYN calculations and the evolved atmosphere approximations.  Furthermore, a smaller Balmer jump ratio is produced in the \dmehard\  evolved-1 model than in either of the 
evolved approximations of the \dmesoft\, as in the RADYN calculations.   A lower Balmer jump ratio in the emergent radiative
flux spectrum in the evolved-1 approximation of the \dmehard\  is due to the combination of the lower
optical depth in the CC ($\tau_{\rm{4170}}(\rm{CC})\sim0.6$ in the \dmehard\  vs. $\tau_{\rm{4170}}(\rm{CC})\sim0.8$ in the \dmesoft), and the higher
temperatures (and thus with larger ambient electron density and larger continuum
emissivity) in the stationary flare layers in comparison to the evolved-2 time in the \dmesoft\ \citep[see][]{Kowalski2016}.
The evolved-2 time of the \dmehard\  has a smaller Balmer jump ratio than the evolved-1 time due to a very large change in optical depth from $\tau \sim3$ to $\sim7$ in the CC at $\lambda=3500$ \AA, resulting in a net decrease in emergent intensity from the atmosphere.  At the evolved-2 time, nearly $\sim25$\% of the emergent blue $\lambda=4170$ \AA\  continuum intensity originates from the heated stationary flare layers with very high electron density ($\sim 3\times10^{15}$ cm$^{-3}$) even though the optical depth in the CC, $\tau_{4170}(\rm{CC})$, is greater than 1.

\clearpage
\floattable
%\begin{turnpage}
\begin{deluxetable}{ccccccccccccccc}
\rotate
\tabletypesize{\scriptsize}
\tablewidth{8.5in}
\tablecaption{Approximate Model Atmospheres:  Continuum optical depth and emergent intensity}
\tablehead{
\colhead{RADYN Model (time)} &  \colhead{log $m_{\rm{ref}}$/g cm$^{-2}$}  & \colhead{$T_{\rm{ref}}$ [K] ($T_{\rm{minCC}}$ [K])} & \multicolumn{2}{c}{ $\tau_{\rm{3500\AA}}(\rm{CC})$} & \multicolumn{2}{c}{$\tau_{\rm{4170\AA}}(\rm{CC})$} &  \multicolumn{2}{c}{$\tau_{\rm{2826\AA}}(\rm{CC})$}  & \multicolumn{2}{c}{$\tau_{\rm{6010\AA}}(\rm{CC})$} & \multicolumn{2}{c}{$F_{3500{\rm{\AA}}}/F_{4170{\rm{\AA}}}$}  & \multicolumn{2}{c}{max $n_e / 10^{14}$ cm$^{-3}$}  \\
 -- & -- & -- & \colhead{approx} & \colhead{ RAD} & \colhead{approx} & \colhead{RAD} & \colhead{approx} & \colhead{RAD} & \colhead{approx} &\colhead{RAD} & \colhead{approx} & \colhead{RAD} &\colhead{approx} & \colhead{RAD}  } 

\startdata
% I used findtau.pro for RADYN optical depth at bottom of CC.  
\solsoft\ evolved-1 (3.97~s)	       &  -2.75  ($t=1.2$~s)   &      9500 (10,000)  &  0.10 &  0.10 & 0.01 & 0.01 & 0.05 & 0.06 & 0.03 & 0.03 & 9.1 & 8.2 & 4.9 & 5.3\\
\solsoft\ evolved-2 (5~s)	       &  -2.75  ($t=1.2$~s)   &      9500 (9500) &  0.15 &   0.18   & 0.02   & 0.02     & 0.09   & 0.10     & 0.04  & 0.04  & 8.6 & 8.0 & 7.2 & 6.7 \\
\hline
\dmesoft\ evolved-1 (1.6~s)    & -2.10  ($t=0.4$~s)   &     11,100 (12,600) &  2.4 &  2.0  &   0.39      &  0.4     & 1.3     & 1.1      & 1.0   & 0.9   & 2.3 &  2.6 &  40 & 50 \\
\dmesoft\ evolved-2 (2.2~s)    & -2.10  ($t=0.4$~s)   &      11,100 (12,100) &  4.9 &  5.4 &   0.75      &  0.8   & 2.7     & 3.0      & 2.0   & 2.2  & 1.94 & 2.1 & 54& 62\\
\hline
\dmehard\ evolved-1 (2.2~s)  & -2.04   ($t=0.4$~s)    &   11,800 (13,300)  & 2.89 &  3.4  &  0.52     & 0.6    &  1.6    &  1.9     &  1.4  & 1.5  & 1.78 & 1.8 & 51 & 51 \\
\dmehard\ evolved-2 (--) & -2.04 ($t=0.4$~s)   &    11,800 (12,800) &  6.6 & -- & 1.1 & -- & 3.7 & -- & 2.9 & -- & 1.7 & -- & 67  & -- \\
\enddata
\tablecomments{The optical depth values at several continuum windows ($\lambda=3500, 4170, 2826, 6010$ \AA) are calculated at the bottom of the CC where the downflowing speed decreases below 5 km s$^{-1}$.  A value of $\mu=0.95$ was used for the optical depth calculations in the approximate model (``approx''') to compare to a $\mu$ value in the RADYN (``RAD'') calculation.  The values of FcolorB for the F13 models calculated with RADYN were obtained from
\citet{Kowalski2016}.  For the evolved-1 approximations, we compare to the RADYN simulations at $t=3.97$~s in the \solsoft\ RADYN model, $t=1.6$~s in the \dmesoft\ RADYN model, and $t=2.2$~s in the
\dmehard\  RADYN model. The evolved-2 approximation of the \dmesoft\ model is compared to $t=2.2$~s in the RADYN calculation.  To obtain $T_{\rm{min,CC}}$ for the lower beam flux density evolved atmospheres, we add 500 K to $T_{\rm{ref}}$ at evolved-1 times and no increase to $T_{\rm{ref}}$ at evolved-2 times. 
To obtain $T_{\rm{minCC}}$ for the high beam flux density evolved atmospheres, we add 1500 K to $T_{\rm{ref}}$ for the evolved-1 approximations and 1000 K to $T_{\rm{ref}}$ for the evolved-2 approximations. The values of $F_{3500{\rm{\AA}}}/F_{4170{\rm{\AA}}}$ indicate the Balmer jump ratios, FcolorB, of the emergent radiative flux density spectrum in units of erg cm$^{-2}$ s$^{-1}$ \AA$^{-1}$. }
\label{table:ccparams}
\end{deluxetable}
%\end{turnpage}

\clearpage

\section{Discussion and Application} \label{sec:application}

Our prescription for parameterizing RHD flare models is an alternative
modeling approach to traditional phenomenological/semi-empirical, static flare modeling that varies atmospheric
parameters through a large possible range \citep{Machado1980,
  Cram1982, Avrett1986, Machado1989, Mauas1990, Christian2003,
  Schmidt2012, Fuhrmeister2010, Rubio2017, Kuridze2017} or static synthetic, beam-heated models
\citep[e.g.,][]{Ricchiazzi1983, HF94}.  Many of these
models are currently
widely used \citep[e.g.,][]{Heinzel2012, Trottet2015, Kleint2016, Simoes2017}.
 When velocity or the
position of the flare transition region is modified in phenomenological
models, the gas density must also change and is not correctly given by hydrostatic equilibrium.  In our approximate models, we
employ density stratifications that self-consistently result from
pressure and
velocity 
gradients in the atmosphere.
The evolved-1 and evolved-2 approximate model atmospheres can be used to explore large grids of model
predictions for the NUV and optical continuum radiation for values of
$m_{\rm{ref}}$, $T_{\rm{ref}}$, and log $g$; an
interesting parameter
space can then be investigated with RHD
simulations for  
 NLTE predictions of the emission line profiles with accurate treatments of
broadening, non-equilibrium ionization/excitation, and
backwarming of the photosphere/upper photosphere. 

There are several assumptions made in our prescription that
limit the accuracy of the continuum predictions in the approximate evolved-1 and
evolved-2 atmospheres.  

\begin{itemize}

\item First, one must assume a
temperature stratification of the stationary flare
layers to obtain the emergent intensity.  
We assume either $\Delta T = -3000$ K (for hard beams) or $\Delta T = -5000$ K (for
soft beams) over the height range of the stationary flare layers.  To
approximate 
the temperature evolution of the stationary flare layers from the
early to evolved times, we assume either no increase occurs or
values of $\Delta T = +500$ K, 1000 K, or 1500 K occurs as in the
RADYN simulations.  The precise values depend on
the flux density, hardness, and evolution of the electron beam energy
deposition.  

For variable
beam parameters over short times, such as the inferred soft-hard-soft
power-law index variation \citep{Grigis2004}, the values of
$m_{\rm{ref}}$ and $T_{\rm{ref}}$ may change significantly and RADYN
simulations are required.

\item Second,
our prescription assumes LTE, which is
satisfactory for the optical and NUV continuum wavelength predictions
for CCs that become sufficiently dense. 
At the evolved-1 time of the \solsoft\ model, the assumption of LTE results in a small
error in the $n=2$ opacity in the uppermost 1 km of the CC.  Using the
snapshot calculated at $t=3.97$~s by the RH code
\citep{Uitenbroek2001} and the contribution function analysis from
\citet{Kowalski2017A}, we find that approximately 10\% of the emergent
NUV continuum intensity originates from the top of the CC where the
NLTE population density of $n=2$ departs by more than 1.7 from LTE; the $n \ge 3$ populations exist at their LTE values. 

\item Third, our approximate model atmospheres do not include a
  parameterization of heating
in the upper photosphere, such as from radiative backwarming due to
Balmer and Paschen continuum photons
\citep{Allred2006}\footnote{This backwarming is the increase in the
  photospheric/upper photospheric temperature
  that results from a radiative flux divergence in the
  internal energy equation \citep[see][]{Allred2015}.  The radiative
  flux is calculated from integrating the solution of the equation of
  radiative transfer.  In the upper photosphere, the increase in
  ionization and temperature is caused by
  Balmer and Paschen continuum photons (from the CC and stationary flare layers)
  that heat the plasma is due to the H-minus bound-free opacity (in
  LTE) and to a lesser degree the hydrogen
  Balmer and Paschen bound-free opacities (in NLTE).}.  Low to moderate
heating ($\Delta T \lesssim 1000$ K) of the upper solar photosphere by
backwarming does not produce significant
continuum radiation at NUV wavelengths \citep{Kleint2016} but it does
affect the Balmer jump ratio due to the increase of H$^-$ emissivity
from the upper photosphere at
red optical wavelengths in the \solsoft\ RADYN model \citep[Appendix A of][]{Kowalski2017A}.  The
upper photospheric heating is not included in our approximations,
which results in larger Balmer jump ratios than with
backwarming included in the RADYN calculations (see Table \ref{table:ccparams}).
In the \dmehard\ and \dmesoft\ models, radiation from upper photospheric
heating does not contribute to the emergent radiation at any wavelength because the 
stationary flare layers and the CC are optically thick at NUV and
optical continuum wavelengths. 
 
\item Fourth, approximations for atmospheres with a surface gravity
that differs from log $g=4.75$ and log $g=4.44$
do not include the modifications to the density stratification in the
CC from large deviations of the velocity field compared to the
5F11 template stratification (see Section \ref{sec:superflares}), and they do not include modifications of
the density stratification of the  
stationary flare layers (see Section \ref{sec:highg}).  More templates for a range of heating
scenarios and surface gravities can be easily included in 
future work if needed.

\item Finally, flares consist of heating and cooling loops; multithread
modeling with complete RHD calculations \citep{Warren2006, Fatima2016,
  Osten2016} are required for a direct comparison to spatially
unresolved observations.  The Balmer jump ratios from multithread modeling are larger
than the extreme values attained at the evolved-1 or evolved-2 times. In
future work, an approximate decay phase parameterization can be developed for 
a course superposition of an early-time (from RADYN), an evolved time
(an evolved-1 or evolved-2 approximation), and a decay-time
(approximation) as a multithread model
\citep[e.g. following][]{Kowalski2017B}
to compare directly to observations. 

\end{itemize}

The agreement between the Balmer jump ratios in the RADYN
simulation and the approximate evolved models (Table
\ref{table:ccparams}) justifies these assumptions and approximations.
The prescription for estimating the optical depth and emergent
continuum intensity in a flare atmosphere consisting of 
a cooled CC and heated stationary flare layers at
pre-flare chromospheric heights
has important applications for interpreting and understanding
the NUV and optical continuum radiation in solar and stellar flares.  
In this section, we present several applications of our approximate model
atmospheres:  constraining flare heating scenarios that produce intermediate
Balmer jump ratios
as observed in some impulsive phase dMe flare spectra (Section \ref{sec:bj}), understanding the role of surface
gravity on the CC density evolution (Section \ref{sec:superflares}, \ref{sec:highg}),  determining the threshold
for hot blackbody radiation in the impulsive phase of dMe flares (Section
\ref{sec:bb}), understanding the interflare variation and relationship
among peak flare colors (Section \ref{sec:relation}), 
and constraining models of solar flares with IRIS data of the NUV
($\lambda \sim 2830$ \AA)
flare continuum intensity (Section \ref{sec:iris}).

\subsection{Application to dMe Flares:  Intermediate Balmer Jump
  Ratios} \label{sec:bj}

The Balmer jump ratio values from the approximate evolved models
(Table \ref{table:ccparams}) can be used to distinguish flare heating scenarios
with NUV and optical continuum radiation formed from material at
$T\sim10,000$ K over low (e.g., as in the
5F11 model) and high (e.g., as in the F13 models)
continuum optical depth.  As an application to dMe flare spectra, we derive the value of $m_{\rm{ref}}$ 
to be achieved at an early time in an RHD flare simulation to produce an
intermediate Balmer jump ratio between FcolorB $\lesssim 2$ (as in
the F13 models) and FcolorB $\gtrsim8$ (as in the 5F11 model).
Values in the range of FcolorB $=3-4$ have been observed at the peak times
of several flares in EV Lac \citep{Kowalski2013} and YZ CMi
\citep{Kowalski2016}.  The hybrid-type (HF) and
gradual-type flare (GF) events
classified in \citet{Kowalski2013} and \citet{Kowalski2016} always exhibit these
intermediate values of the Balmer jump ratio in the impulsive phase; some
impulsive-type flare (IF) events also exhibit these values at peak times
\citep{Kowalski2016}, while the gradual decay phases of all types of flare
events can be characterized by intermediate Balmer jump ratios $\gtrsim
2.75$ \citep{Kowalski2013, Kowalski2016}.  Thus, the intermediate
Balmer jump ratio values are an important observed phenomenon to reproduce with RHD models.  

\begin{figure}[h!]
\plotone{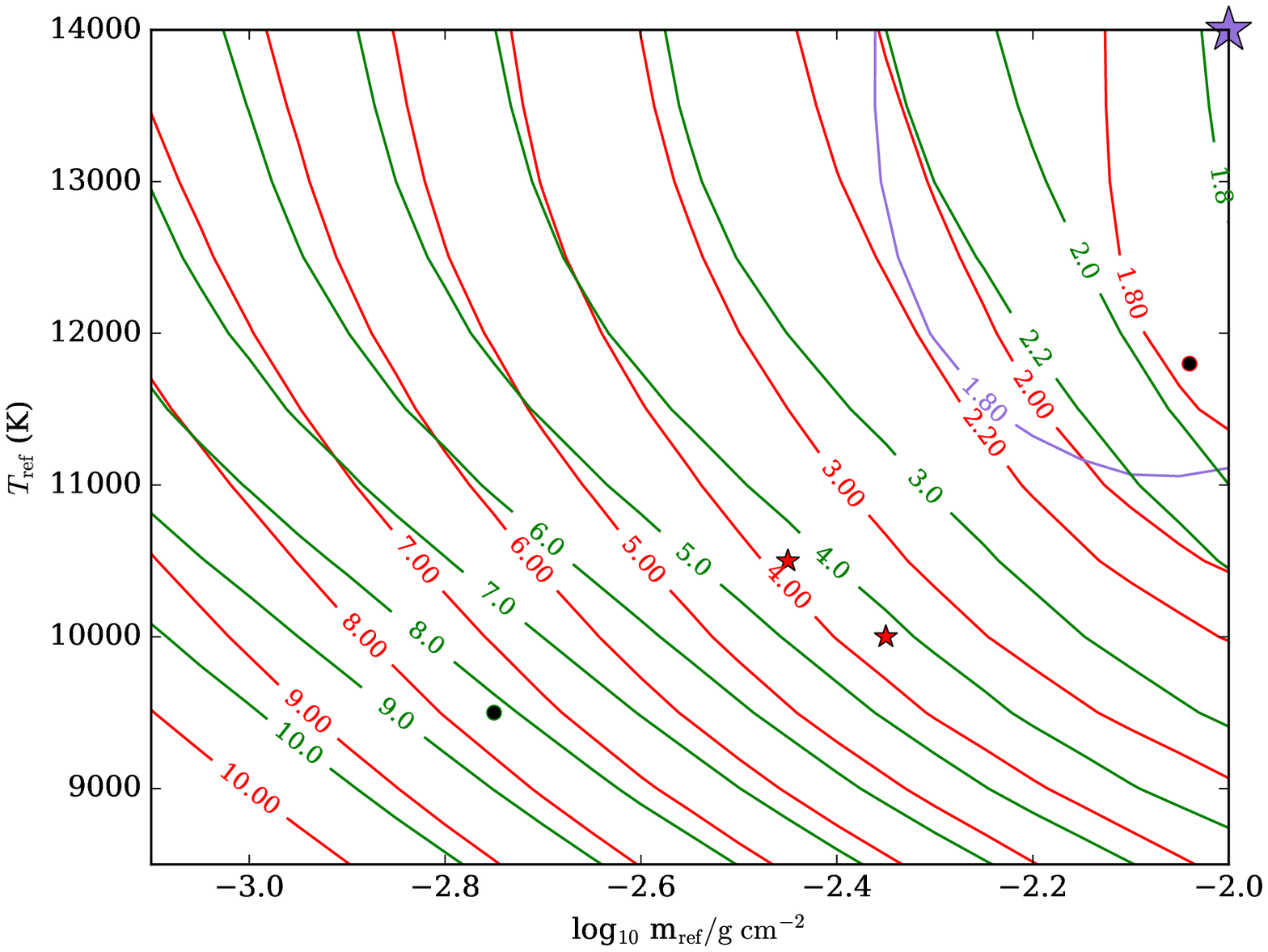}
\caption{Contours of the Balmer jump ratio, $F_{3500}/F_{4170}$,
  denoted as FcolorB in \citet{Kowalski2016}.  Red contours correspond
to the dMe model approximations ($x_{\rm{maxCC}}=15$ km), and green
contours correspond to the solar model approximations
($x_{\rm{maxCC}}=30$ km).  We use the evolved-1 approximation with
$T_{\rm{minCC}}= T_{\rm{ref}}+1500$ K; the heating in the stationary flare
layers is approximated for a hard electron beam distribution for all
calculations (see text).  Red stars indicate 
example dMe flare evolved-1 atmospheres that produce an intermediate
Balmer jump ratio of FcolorB$=3.6-3.8$.   The
black circles indicate
the Balmer jump ratios from the approximate models at the evolved-1
times for the \solsoft\ model (Figure \ref{fig:5F11CC})
and the \dmehard\ model
(Figure \ref{fig:F13_evolvedA}).  
The purple contour shows the range of parameters that produce a Balmer
jump ratio of FcolorB $=1.8$ in the evolved-2 approximation for the
dMe surface gravity.  For the dMe gravity, the
lowest Balmer jump ratio in this parameter space for the evolved-2 approximation is 1.6 which is indicated
by the purple star; the lowest Balmer jump ratio in the parameter
space for the evolved-1 approximation is 1.7.       }
  \label{fig:grid}
\end{figure}

In Figure \ref{fig:grid}, we show contours of the Balmer jump ratio, FcolorB, for a range of
$T_{\rm{ref}}$ and log $m_{\rm{ref}}$, calculated in intervals of 500 K
and 0.05, respectively, with our approximate models at the evolved-1
times. The Balmer jump ratio has been calculated 
for the dMe surface gravity in red contours.  For large values of $m_{\rm{ref}}$, the Balmer jump ratio
decreases for increasing $T_{\rm{ref}}$ at constant $m_{\rm{ref}}$
and for increasing $m_{\rm{ref}}$ at constant $T_{\rm{ref}}$. 

 The evolved-1 times represent a
conservative estimate to the minimum Balmer jump ratio attained in an
RHD model.  The
Balmer jump ratio becomes lower at the evolved-2 time because the 
CC attains a higher column mass by a factor of $\sim1.5$; the minimum
Balmer jump ratio depends on the poorly constrained duration of high
beam flux heating in a flare loop.    We show a purple contour
in Figure \ref{fig:grid} for the value of FcolorB $=1.8$ to illustrate
the difference between the evolved-1 and evolved-2 times.  In the upper right corner of Figure \ref{fig:grid}, the lowest
Balmer jump ratio is 1.7 in the evolved-1 approximation and 1.6 in the
evolved-2 approximation.

Red star symbols on this figure show that an
intermediate Balmer jump ratio of FcolorB$=3.6-3.8$ at the peak of a dMe flare can be produced by an
atmosphere with $T_{\rm{ref}}=10^4$ K and log
$m_{\rm{ref}}=-2.35$ or with $T_{\rm{ref}}=10,500$ K and
$m_{\rm{ref}}=-2.45$.  At these values, the optical depth  (at
$\mu=0.95$)\footnote{The $\mu$ value here is chosen arbitrarily as
  0.95 to represent the optical depth and is a standard $\mu$ value
  used in radiative transfer codes.  The optical depth at any other
  $\mu$ value for a plane-parallel atmosphere can be obtained by
  multiplying the optical depth at $\mu=0.95$ by $0.95 / \mu$.}
within
the CC at $\lambda=4170$ \AA\ is $\sim0.1$ and at $\lambda=3500$ \AA\
the optical depth is
between $0.6-0.9$:
because $\tau_{3500}$ in the CC is near one, a significant amount of outgoing Balmer
continuum radiation from the stationary flare layers is attenuated,
and the Balmer jump ratio in an emergent spectrum is smaller than the
Balmer jump ratio in an emergent spectrum 
formed at $T\sim10^4$ K over low continuum optical depth.  The maximum value of the electron
density that results in a CC with log $m_{\rm{ref}}=-2.35$ is $n_e\sim2\times 10^{15}$ cm$^{-3}$.  Thus, we 
expect broad hydrogen lines from CCs that produce the
intermediate Balmer jump ratio, but NLTE modeling with an
accurate broadening treatment is required for detailed line shapes \citep{Kowalski2017B}.  
Results from the approximate evolved-1 atmospheres will be used to
compare to new RHD models with high
low-energy cutoff values in a future work on flares with intermediate
Balmer jump ratios observed in the dM4e star GJ 1243 (Kowalski et al. 2018A, in prep).

\subsection{Application to Superflares in Rapidly Rotating dG Stars} \label{sec:superflares}

 The approximate model atmospheres can be used to understand how the emergent continuum
spectral properties vary in flares occurring in stars over a range of surface
gravity values.  The heating in the lower
atmosphere in superflares
($E_{\rm{white-light}}\approx10^{35}-10^{36}$ erg) observed by \emph{Kepler} in rapidly rotating, young dG
stars \citep{Maehara2012} is not yet understood. Compared to the largest flares in the present day Sun, do these
superflares result from
larger average energy flux densities in electron beams and/or do they
exhibit larger flare
areas?  
In this section, we explore the Balmer jump ratio expected from lower
surface gravity solar-type stars, though such constraints are
currently not
readily 
available for solar flares or dG superflares.  

Contours of the Balmer jump ratio calculated with the value
of $x_{\rm{maxCC}}=30$ km (for solar surface gravity) are shown as the green contours in Figure \ref{fig:grid}.
All other parameters are kept the same in these evolved-1 calculations
compared to the dMe (red) contours with stationary flare
layer heating estimated for a hard electron beam distribution.  For similar
values of $T_{\rm{ref}}$ and log $m_{\rm{ref}}$, the solar
atmosphere produces larger values of the Balmer jump ratio FcolorB.  This
difference occurs due to the different surface gravitational
acceleration, log $g$, by a factor of two.  For the same electron beam flux density and beam energy distribution 
in a dG and dMe star flare, a similar column mass is heated by the
beam \citep{Allred2006}, and we expect this to produce a similar value of 
$m_{\rm{ref}}$ among atmospheres of different gravity.  The factor of
two larger gravity in a dMe star and the same value of log
$m_{\rm{ref}}$ means that this column mass of material is
compressed into a factor of $\frac{10^{4.75}}{10^{4.44}}=2$ smaller
physical depth range.  The value of $m_{\rm{ref}}$ is an area in 
$\rho(z)$ vs. $z$, which results in a larger value of $\rho_{\rm{maxCC}}$, 
continuum optical depth, and maximum electron density in a CC.

For large beam flux densities near F13, however, the relationship
between gravity and Balmer jump ratio becomes
more complicated than implied by the scaling in Equation
\ref{eq:template}.  
We run a RADYN flare simulation of the solar atmospheric response to a
very high beam flux density, F13, $\delta=3$, $E_c=37$ (hereafter,
\solhard)\footnote{We use a similar starting atmosphere to the
  \solsoft\ model.}, for direct
comparison to the \dmehard\ and \dmesoft\ models.  The energy deposition
lasts for 4.5~s, at which point the coronal temperature exceeds 100
MK, which is the upper limit to the atomic data currently in
RADYN.  The physical
depth range of the CC in the \solhard\ simulation is $x_{\rm{maxCC}} \sim 30$
km, confirming that this parameter is independent of the beam flux density and is
inversely proportional to 
the surface gravitational acceleration.  At $t=0.4$~s, we calculate that log $m_{\rm{ref}} = -1.96$ and
$T_{\rm{ref}}=11,650$ K using the algorithm in Section \ref{sec:method};
note that these values are similar to the values that result from
flare heating in the \dmehard\ RADYN model (Table \ref{table:ccparams}). 
The approximate evolved-2 atmosphere prescriptions predict that 
the maximum electron density in the CC is $4.3 \times 10^{15}$
cm$^{-3}$, the optical depth at $\lambda=4170$ \AA\ at the bottom of
the CC is 0.9 and the value of FcolorB is 1.7.  In the \solhard\ RADYN calculation, the maximum electron density in
the CC is much larger  $\sim7 \times 10^{15}$ cm$^{-3}$ compared to
this prediction. We inspect
the density stratification of the \solhard\ model compared to the template
density stratification obtained from the \solsoft\ at the evolved-1
time (Figure \ref{fig:5F11CC}). There is a significant difference in
the density stratification compared to the \solsoft\
density template stratification.  In the
\solhard\ model,
much larger downflow speeds of 200 km s$^{-1}$ occur, compared
to $\sim50-60$ km s$^{-1}$ in the \solsoft\ and $90-100$ kms$^{-1}$ in
the \dmehard\ and \dmesoft\ models.

From Equation \ref{eq:template}, the
larger gravity of the dMe atmosphere results in more atmospheric compression
over height; a similar value of $m_{\rm{ref}}$ is expected to
produce a factor of $\sim2$ lower maximum density in a solar model atmosphere
for the same beam flux density of F13.
However, the RADYN \dmehard\ and \solhard\ simulations
have roughly the same maximum density ($\rho_{\rm{maxCC}}$) in the CCs. 
 The analytic calculations from
\citet{Fisher1989} show that the pre-flare chromospheric density just below the flare
transition region is inversely related to the maximum downflow speed.  
The \solhard\ and \dmehard\ models exhibit a similar value of
$m_{\rm{ref}}$ (to within 20\%), but the initial uncompressed density just
below the flare transition region is smaller in the solar atmosphere
(also by a factor of two)
which results in $\sim$two times larger maximum downflow speeds ($\sim200$ km
s$^{-1}$) in the solar CC.  The product of the maximum downflow
speed ($v$) and preflare gas density ($\rho$) below the flare transition
region 
gives a similar initial mass flux density (in units of $g$ cm$^{-2}$
s$^{-1}$) in the two F13 models.  Using the template gas density
stratification from the \solsoft\ model implicitly assumes that the
mass flux density (which determines the amount of gas compression and thus $\rho_{\rm{maxCC}}$) is controlled by the ratio
of preflare gas density values below the flare transition region and
is not influenced strongly by much larger or smaller initial
downflow speeds.

The solar 
contours in Figure \ref{fig:grid} therefore underestimate the values of the actual
Balmer jump ratios of the emergent spectra.  The RADYN
simulation shows that the Balmer jump ratio at the evolved-1 time
(2.2~s) is 1.7 compared to 1.9 in Figure \ref{fig:grid};  at the
evolved-2 time (2.8~s), the Balmer jump ratio attains a value as low
as 1.5 in RADYN,
compared to the evolved-2 approximation of 1.7.  Furthermore, the Balmer jump ratio
in the \solhard\ model is lower than our predicted Balmer jump ratio at the evolved-2 time of
the \dmehard\ model (Table \ref{table:ccparams}).
 By adjusting\footnote{We multiply the \solsoft\ density stratification template
   by an exponential function that reflects the velocity
   difference of 200 km s$^{-1}$ and 50 km s$^{-1}$, and we re-solve Equation
   \ref{eq:template}. } the \solsoft\ density stratification template using the difference in
downflow speeds between the \solhard\ and \solsoft\ models, we predict an electron density of
$6.6 \times 10^{15}$ cm$^{-3}$ and a larger blue continuum optical depth at
the bottom of the CC ($\tau_{4170}(\rm{CC}) \sim 1.1$) for log $m_{\rm{ref}}=-1.96$.  These are closer
to the values in the \solhard\ RADYN simulation ($7 \times 10^{15}$ cm$^{-3}$
and 1.2, respectively).  The template from the \solsoft\ model is accurate for predictions of the maximum electron
density in the CC for a limited range of
downflow speeds to within a factor of the \solsoft\ maximum downflow speeds.

Simultaneous X-ray and optical spectra of
dG superflares would determine if 
high flux electron beam flux densities are generated in young solar-like
stars.  Notably, large
energy flux densities between F12 and F13 have been inferred in bright solar flare
kernels \citep{Neidig1993, Krucker2011, Sharykin2017}, and we may expect values of the Balmer
jump ratio as low as FcolorB $\lesssim2$ in spatially resolved, solar
flare spectra as well as in spectra of dG superflares.

\subsection{Application to dMe Flares:  Hot Blackody-like Radiation} \label{sec:bb}
In the impulsive phase of some dMe flares, an energetically important observed spectral property is a color temperature of
$T\sim9000-14,000$ K in the
blue and red optical wavelength range \citep{HP91, Zhilyaev2007,
  Fuhrmeister2008, Kowalski2013}.   The emergent flux spectra with a
color temperature of $T\sim9000-14,000$ K  also exhibit small Balmer
jump ratios \citep[FcolorB$<$2; cf Figure 12 of][]{Kowalski2016}.  
We calculate the FcolorR continuum flux ratio, $F_{4170}/F_{6010}$,
which is a proxy of the blue-to-red optical color temperature 
\citep{Kowalski2016} for our approximate model atmospheres.  Contours are shown in Figure \ref{fig:FcolorR}
for the 
evolved-1 atmosphere approximations with stationary flare layers
heated by a hard beam ($\delta \sim 3$).  We define ``hot'' color temperatures as 
$T_{\rm{FcolorR}} \ge 8500$ K, corresponding to FcolorR $=1.7$ and the
thick red contour.   This thick contour is the threshold that we establish for producing
hot blackbody-like radiation for a high electron beam flux with a hard
power-law distribution.  We do not address here the property of some
flares exhibiting
blue optical continua ($\lambda=4000-4800$ \AA) with a larger color
temperature (by $\Delta T \sim2000$ K) than indicated by $T_{\rm{FcolorR}}$
\citep{Kowalski2013, Kowalski2016}; our approximate models here generally produce a blue
color temperature lower than $T_{\rm{FcolorR}}$ by several hundred K.

The \dmehard\ and \dmesoft\ models reproduce hot color temperatures at their evolved-1 and
evolved-2 times and
possibly explain this interesting spectral phenomenon
\citep{Kowalski2013, Kowalski2016}.  
 The location of the evolved-1 time of the \dmehard\ 
 model is shown in Figure \ref{fig:FcolorR}
as a black circle.  
 Detailed modeling of optical spectra with the RH code shows
that extremely broad Balmer lines result from
the high charge density ($n_e\sim 5\times10^{15}$ cm$^{-3}$) in the CC
\citep{Kowalski2017B}.  Also shown in Figure \ref{fig:FcolorR} are
contours of the maximum (LTE)
electron density achieved in the evolved-1 approximations.  
We thus expect very high electron densities in the CCs for the range of FcolorR values
that are consistent with the hot color temperature observations.  A large beam flux density is expected to produce a
strong return current electric field and beam instabilities
\citep{Holman2012,Li2014}, which were not included in any of the
RADYN simulations or approximations in this work.   Also, the coronal magnetic field energy that is converted to
kinetic energy during reconnection must be $B\sim$1.5 kG
\citep{Kowalski2015}.  These issues place constraints on the
highest beam flux density that is possible in the atmospheres of
stars.  
The values of $m_{\rm{ref}}$ and $T_{\rm{ref}}$ from any future RHD
model with lower beam flux density than F13 can be placed on Figure
\ref{fig:FcolorR} (or a similar figure for
soft $\delta \gtrsim 4$ electron beams; not shown) and compared to the
thick red contour
to determine if the model is expected to produce hot blackbody continuum radiation at $\lambda > 4000$ \AA.

\subsubsection{Implications for High Gravity dMe Flare Models} \label{sec:highg}
\begin{figure}[h!]
\plotone{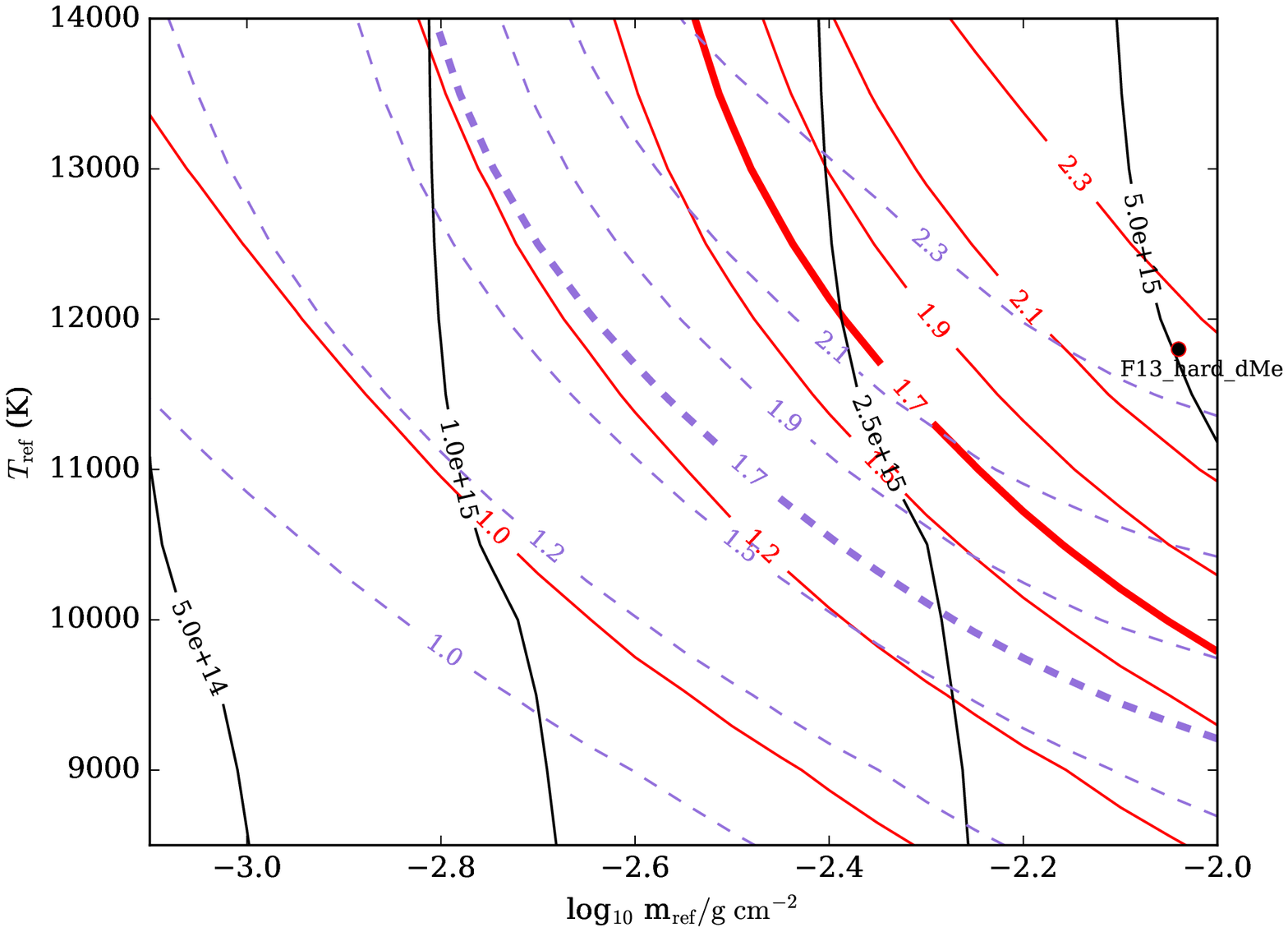}
\caption{Contours of the blue  ($\lambda=4170$ \AA) continuum flux divided by the red
  ($\lambda=6010$ \AA)
  continuum flux, referred to as FcolorR in \citet{Kowalski2016}, for
  the evolved-1 approximations (with hard beam heating).  We set $T_{\rm{min,CC}}= T_{\rm{ref}}+1500$ K for all calculations.  Red contours correspond
to the dMe model approximations ($x_{\rm{max,CC}}=15$ km).
The black circles indicates
the ratios from the approximate model of the \dmehard\ RADYN
simulation at the evolved-1 time.  
The color temperatures, $T_{\rm{FcolorR}}$, corresponding to these
contours are $T_{\rm{FcolorR}}=(5800, 6500, 7600, 8500, 9300, 10400, 11600)$ K for
FcolorR $=(1.0, 1.2, 1.5, 1.7, 1.9, 2.1, 2.3)$, respectively.  
 Also shown are contours for the maximum
electron density produced in the evolved-1 models of the CC.  The
maximum electron density is always greater than $2\times10^{15}$
cm$^{-3}$ for the atmospheres that have $m_{\rm{ref}}$ and
$T_{\rm{ref}}$ that also produce hot ($T_{\rm{FcolorR}} \ge 8500$ K; thick red
contour)
color temperature that characterizes the blue-to-red optical continuum.
The dashed purple contours indicate the values
of FcolorR for evolved-1 approximations in a dMe atmosphere with 
$x_{\rm{maxCC}}=8$ km to mimic the compression in an atmosphere with
log $g=5$.  No adjustment for the density
stratification of the stationary flare layers was made compared to the
log $g=4.75$ red contours.  The thick red contour is our determination
for the threshold for forming a hot blackbody flux spectrum at
$\lambda>4000$ \AA\ for hard electron beams.   }
  \label{fig:FcolorR}
\end{figure}

The RADYN simulations of dMe flares that are used to 
produce the contours in Figures \ref{fig:grid} - \ref{fig:FcolorR} 
have log $g=4.75$.  Using the fundamental parameter relationships in \citet{Mann2015} and the
magnitude and distance information from \citet{NLDS}, we calculate that the surface
gravity values for AD Leo (dM3e), YZ CMi (dM4.5e), and Proxima
Centauri (dM5.5e) to be log $g=4.83$, log $g=4.9$, and log $g=5.26$, respectively.  We show purple
contours in Figure
\ref{fig:FcolorR} for the approximate FcolorR values from 
evolved-1 atmospheres with a CC having a physical depth range of
$x_{\rm{maxCC}}=8$ km, which is estimated as the physical depth range
of a CC in an atmosphere with log $g=5$.   We adjust the physical depth
of the stationary flare layers according to the surface gravity
(Appendix A.2) but
no change is made to the density stratification in the stationary
flare layers compared to the log
$g=4.75$ hydrostatic equilibrium stratification.  The purple contours in Figure
\ref{fig:FcolorR} indicate that lower electron beam flux density
values may result in similar valus of $m_{\rm{ref}}$ but larger CC
densities and continuum optical depths in high gravity
M dwarfs.  Changing the surface gravity would also affect the maximum downflow
speed for the same nonthermal beam density (Section \ref{sec:superflares}), and accurate approximations for higher surface gravity 
may require a CC density stratification template from a new
hydrodynamic simulation.  We suggest that RADYN calculations explore higher surface gravity
 values of log $g>4.75$.

A large sample of flux ratio measurements \citep[e.g., with
ULTRACAM;][]{Dhillon2007, Kowalski2016} of flares in stars spanning stars of the
 subtypes dM0e-dM7e with similar values of log $L_{\rm{H}\alpha}$ / $L_{\rm{bol}}$
would test whether surface gravity affects
the appearance of optical color temperatures of $T_{\rm{FcolorR}} \ge$ 8500 K and small Balmer jump
ratios of FcolorB$<$2 in the observed spectra.
 Curiously, the Great Flare of AD Leo, which has a log
$g=4.83$ that is close to the RADYN model atmosphere (as chosen
initially), exhibits
a very small Balmer jump ratio of $1.4-1.8$ and a hot optical flare
blackbody \citep{HP91,HF92, Kowalski2013}.  Thus a larger surface gravity
with a significantly lower beam flux density (and thus a smaller value
of $m_{\rm{ref}}$ for a comparably large value of FcolorR $\gtrsim2$) cannot account for these flare
properties in all active M dwarf stars.

\subsection{Application to dMe Flares: The Interflare Variation of Peak
  Continuum Flux Ratios}  \label{sec:relation}
\begin{figure}[h!]
\plotone{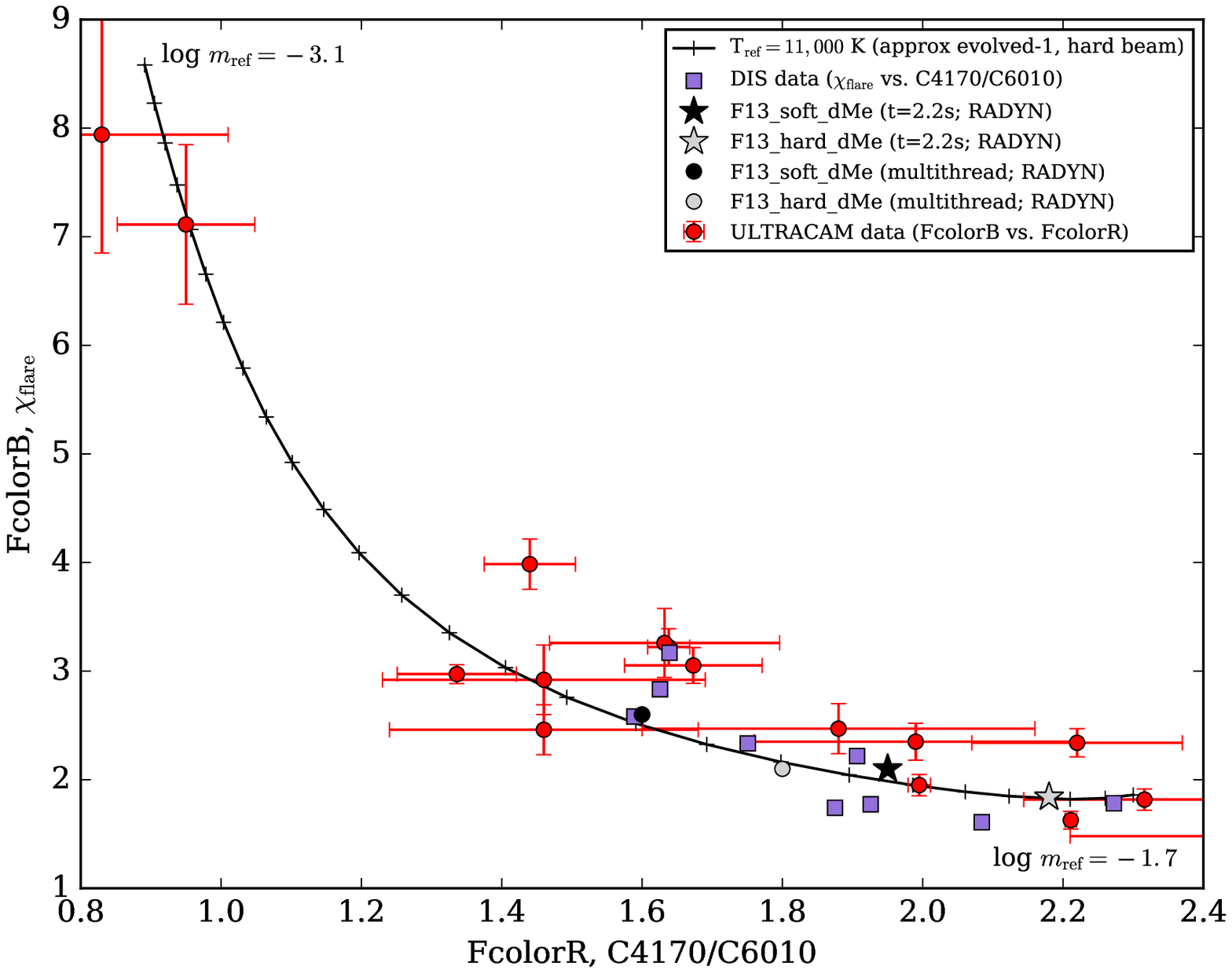}
\caption{The relationship for FcolorB vs. FcolorR with
  $T_{\rm{ref}}=11,000$ K in the approximate evolved-1 models. The
  value of log
  $m_{\rm{ref}}$ is varied from -3.1 to -1.7.  The photometry data from Figure 12 of \citet{Kowalski2016} are shown
  as red circles (from ULTRACAM), and the continuum flux ratios from
  the Apache Point Observatory's ARC 3.5-m/DIS spectra \citep[from][]{Kowalski2013} are shown as purple squares (where
  $\chi_{\rm{flare}}$ is similar to FcolorB and C4170/C6010 is similar
  to FcolorR).  The observed relationship is
  generally reproduced by varying log $m_{\rm{ref}}$.  However, the largest values of FcolorR are
  only produced by an extreme value of log $m_{\rm{ref}} = -1.7$, and
  some flares exhibit a significant offset towards larger values of
  the Balmer jump ratio.  Note that there are
  several flares at very large values of FcolorR $>2.4$ that are not
  shown here;  these secondary flare events may be
  explained by a very high low-energy cutoff electron beam heating
  function \citep{Kowalski2017B}.   The values for the RADYN \dmehard\
  (at the evolved-1 time)
and \dmesoft\ (at the evolved-2 time) are shown as gray and black
stars, respectively;  the average burst (multithread) model for each are shown as
 gray and black circles respectively. }
  \label{fig:relation}
\end{figure}

In Figure \ref{fig:relation}, we show the relationship between FcolorR
and FcolorB predicted from our approximate evolved-1 atmospheres by
varying log $m_{\rm{ref}}$ from -3.1 to -1.7 and keeping
$T_{\rm{ref}}=11,000$ K ($T_{\rm{minCC}}=12,500$ K) constant.  
The flare peak data from \citet{Kowalski2016} shows
that the range of impulsive phase continuum flux ratios may result
from variations of
$m_{\rm{ref}}$ from flare to flare.   The $t=2.2$~s continuum flux ratio values for the
\dmehard\ and \dmesoft\ are shown also as a light gray and black star,
respectively.   \citet{Kowalski2016} proposed
that the differences in the hardness of the electron beam between
these two models may suggest that interflare
peak variation is due to beam hardness variations.  In our
approximations, varying the value of $m_{\rm{ref}}$ represents changing the
electron beam hardness and flux density.  

The approximate model
relationship falls significantly below
the values of FcolorB for 
five of the flares with ULTRACAM data in Figure \ref{fig:relation}.
Larger observed values of FcolorB than in the evolved-1 time
approximations indicate that relatively
more Balmer continuum radiation is produced in the impulsive phase
than is accounted for by variations of $m_{\rm{ref}}$.  This ``missing''
Balmer continuum radiation in the approximate model representation may
result from flare loops that are heated in
the early rise phase and gradually decay through the impulsive phase \citep[e.g., see][]{Warren2006}.  A multithread analysis
of the \dmesoft\ and \dmehard\ RADYN models reveals that superposing all
snapshots of these
models increase FcolorB but also decrease FcolorR \citep[cf Table 1 of][]{Kowalski2017B}.
We show these multithread model values in Figure \ref{fig:relation} as
black and light gray circles.  The multithread models do not
account for the large Balmer jump ratios at high values of FcolorR,
nor do they account for the intermediate Balmer jump ratios of FcolorB
$>3$ (Section \ref{sec:bj}).

In Figure \ref{fig:relation}, we have
extended the
range of log $m_{\rm{ref}}$ to $-1.7$ in order to reproduce some of the largest observed
values of FcolorR.  Such a large value of log $m_{\rm{ref}}$
suggests an electron beam flux density $>$ F13 and thus a very strong
return current electric field.  Neither this extreme value of log $m_{\rm{ref}}=-1.7$
and $T_{\rm{ref}}=11,000$ K nor the largest values in Figure
\ref{fig:grid} (for log $m_{\rm{ref}}=-2$
and $T_{\rm{ref}}=14,000$ K)  produce the smallest values of FcolorB$\sim 1.4-1.5$ observed at the
main peaks of other
dMe flares \citep{HP91} and in some energetic secondary flares \citep{Kowalski2013}.

\subsection{Application to dG Flares:  NUV Continuum Intensity in
  IRIS Spectra} \label{sec:iris}
High spectral resolution and high time resolution observations of the
brightest kernels in the hard X-ray impulsive phase of solar flares exhibit a redshifted component in
the singly ionized chromospheric lines \citep{Graham2015} and in H$\alpha$ \citep{Ichimoto1984} which is
conclusive evidence for the
formation of dense CCs.  Bright NUV continuum intensity is also
predicted from dense CCs \citep{Kowalski2017A}, and high spatial resolution spectra of
bright kernels in solar flares are now
readily available from IRIS to compare to high beam
flux density RHD models.  

The NUV spectra from IRIS include a continuum region at
$\lambda=2825.6-2826.9$ \AA\ (hereafter, ``IRIS NUV'' or
  ``C2826'' as referred to in \citet{Kowalski2017A}) outside of major and minor
emission lines, where the Balmer-continuum enhancement
was first detected by \citet{Heinzel2014}.
The bright emergent continuum intensity in the \solsoft\ RADYN model 
attains a brightness (and becomes brighter) than the observed values of C2826
in the 2014-March-29 X1 solar flare. We calculate the emergent NUV 
continuum intensity from our evolved-1 and evolved-2 model
atmospheres using the values of $T_{\rm{ref}}$ and log
$m_{\rm{ref}}$ in the RADYN calculation.  
  The emergent excess intensity at $\lambda=2826$ \AA\ in the 
  \solsoft\ RADYN model at $t=3.97-5$~s is $5-5.5\times10^6$ \cgs
  \citep{Kowalski2017A} and is consistent with the range of emergent
  intensity predicted by our approximate evolved-1  ($I_{\lambda=2826,
    \mu=0.77}=5.0\times10^6$ \cgs) and evolved-2 ($I_{\lambda=2826,
    \mu=0.77}=6.5\times10^6$ \cgs) atmospheres.
In this case, the evolved-1 and evolved-2 atmospheres provide lower
and upper limits, respectively, to compare
to observational constraints from IRIS.  The approximate values 
of $\tau_{2826}$ at the bottom of the CC are shown in Table
\ref{table:ccparams}; there is satisfactory agreement between our
approximations and the NLTE, NEI calculations with RADYN.  

A lower beam flux density of F11 was used in
\citet{Kowalski2017A} for comparison to the 5F11 model \citep[see
also][for detailed analyses of the F11 model]{Kuridze2015, Kuridze2016}.   The F11 results in faint NUV
continuum intensity that is not consistent with the observations of
the two brightest flaring footpoints at their brightest
observed times in the 2014-Mar-29
flare, thus favoring a higher beam flux density for these kernels.  The low-energy
cutoff that was used was $E_c=25$ keV as inferred from hard X-ray fitting of RHESSI
data.  As well known, this is an upper limit to the low energy cutoff
that can be inferred from the data due to a bright thermal X-ray spectrum at lower
energies.  
We use our
approximate, evolved model atmospheres to determine if a lower beam flux density of F11 and a
lower low-energy cutoff ($E<<25$ keV) reproduces the observed NUV continuum
intensity in the two brightest flare kernels in the 2014-Mar-29 X1 flare.  We use the RADYN code to
simulate the atmospheric response to an F11 beam flux density with $E_c=15$ keV
and $\delta=3.6$, which are also consistent with the hard X-ray
observations \citep{Battaglia2015}.  This value of $\delta$ is harder
than the F11 model with $\delta=4.2$ presented in
\citet{Kowalski2017A}.  Following the Fe II LTE line analysis in \citet{Kowalski2017A},
we find that the harder F11 model with the lower low-energy cutoff produces two Fe II 2814.45 emission line components
that are consistent with the formation of a CC and stationary flare
layers. The F11 with $E_c=25$ keV in \citet{Kowalski2017A} does not
produce a red shifted emission component in Fe II.
In this new RADYN model, we obtain the values of 
log $m_{\rm{ref}} = -3.08$ and $T_{\rm{ref}} = 8900$ K at the early
time of $t=1.2$~s.  The
approximate emergent intensity at the evolved-1 time (with the soft beam heating
approximation in the stationary flare layers) matches the
value of $10^6$ \cgs\ in the RADYN simulation. The
approximate model at the evolved-2 time (with the hard beam heating
approximation in the stationary flare layers) predicts an
emergent continuum NUV intensity of 
$1.4\times10^6$ \cgs\  which is nearly a factor of two below the observed
excess C2826 values\footnote{The emergent continuum intensity from our approximate
  models most closely resemble the \emph{excess continuum intensity}, $I_{\rm{flare}}-I_{\rm{pre}}$, because the radiation field from
  the photosphere is not included in these LTE approximations (the
  approximate evolved models do not include the radiation from a heated
  upper photosphere).  } of $\gtrsim 2.2\times10^6$ \cgs\ in the two
brightest footpoints BFP1 and BFP2 in the 2014-March-29 X1 flare.
Therefore, a F11 flux density cannot
reproduce the brightest NUV continuum intensity observed in this flare;
this heating scenario is an insufficient model for 
understanding the atmospheric processes in the brightest
continuum-emitting kernels.  However, representative values of C2826
for this flare are $<1.5\times 10^{6}$ \cgs\ \citep{Kleint2016} and
these fainter flaring pixels may be explained by lower heating rates. 
Our new F11
RADYN flare model and a new 2F11 RADYN flare model (also with $E_c=15$ and $\delta=3.6$) will be
discussed in more detail in comparison to observations of the hydrogen line
broadening in a future paper (Kowalski et al. 2018B, in prep).

\section{Summary and Conclusions}

We have developed a prescription to predict the approximate values of the NUV and optical
continuum optical depth, the
emergent continuum intensity, the continuum flux ratios, 
and the maximum electron density attained
in flare atmospheres exhibiting an evolved, cooling compression above stationary
heated layers with $T\approx 10^4$ K.
The prescription depends on specifying only two parameters besides the
gravity of the star:
$T_{\rm{ref}}$ and log $m_{\rm{ref}}$, which can be readily obtained
at early times in radiative-hydrodynamic simulations such as with the
RADYN code.  
The approximate, evolved atmospheres provide interesting electron beam 
parameter space selection ($\delta$, $E_c$, flux density) for computationally intensive RADYN and
Flarix \citep{flarix}
simulations of the non-LTE, non-equilibrium ionization/excitation
hydrogen Balmer line and singly ionized chromospheric line
profiles. Our analysis of $m_{\rm{ref}}$ and $T_{\rm{ref}}$ can also be applied to future 3D
flare models that can resolve the large
pressure gradients that drive these chromospheric condensations.  
3D NLTE RHD models will be more computationally expensive than the
current 1D NLTE RHD simulations, and a selective range of electron beam
parameters will be necessary.   Given a CC density stratification
template, our prescription and analysis can be applied to any flare heating
scenario that produces two flare layers at pre-flare chromospheric
heights.

The approximate models of the evolved states of a CC have been used to
determine
the values of $T_{\rm{ref}}$ and log $m_{\rm{ref}}$ that
produce blue optical continuum radiation ($\lambda=4170$ \AA) formed
in two, $T\sim10^4$ K flare layers with a large optical depth
($\tau_{4170} = 0.4-1.2$), an intermediate optical depth
($\tau_{4170}\sim 0.15$), and a low optical depth ($\tau_{4170} <<
0.1$).  Very high
beam energy deposition rates as in the F13 models produce
$m_{\rm{ref}} \sim 0.01$ g cm$^{-2}$ and a large optical depth
($\tau_{4170} = 0.4-1.2$) in the $T\sim10,000$ K material in the CC, which 
results in a larger optical depth and smaller physical depth range at
NUV and red wavelengths due to the wavelength dependence of hydrogen
b-f opacity.  A large optical depth produces an
emergent continuum spectrum with a color temperature of $T\sim10,000$ K and a
small Balmer jump ratio as observed in the impulsive phase of dMe
flares \citep{Kowalski2013}.  We have determined a critical threshold
contour of $T_{\rm{ref}}$ and log
$m_{\rm{ref}}$
(Figure \ref{fig:FcolorR}) for producing the hot blackbody radiation.
This threshold 
can be compared to values obtained from lower beam flux density simulations than
F13, which results in a very strong return current electric field (to
be included in the energy loss in the electron beam in a future work) and
requires strong magnetic fields in the corona. 
 Our prescription predicts that high ambient electron
densities of $n_e>2\times10^{15}$ cm$^{-3}$ in the CC are produced
for all significantly large values greater than log $m_{\rm{ref}}=-2.4$.

 Our approximations are accurate enough to distinguish between
flare atmospheres that result in a large Balmer jump ratio (low
optical depth at all continuum wavelengths) and a smaller Balmer jump
ratio (intermediate or high optical depth at all continuum
wavelengths).  Observations of the peak phases of dMe flares exhibit a
range of properties, and variation of $m_{\rm{ref}}$ in our approximate model atmospheres cover
most of the observed relationship between the Balmer jump ratio
and blue-to-red optical color temperature (Figure \ref{fig:relation}).
Some dMe flares exhibit intermediate Balmer jump ratios at the
peak of the impulsive phase.  Our prescription predicts that
electron beam heating resulting in log $m_{\rm{ref}}
\sim -2.35$ to $-2.45$ at early times would produce an
intermediate Balmer jump ratio in the evolved states of the atmosphere.  For these evolved model atmospheres,
$\tau_{3500}\sim 1$ and $\tau_{4170} \sim 0.15$ at the bottom of the CCs.  We expect that lower beam
energy 
flux densities than F13 can produce such emergent flux spectra with
intermediate Balmer jump ratios, which are also observed in the
gradual decay phase of dMe flares.

The CC density and emergent continuum properties depend
on stellar surface gravity in a complex way.  Generally, larger surface gravity stars
produce denser CCs and larger continuum optical depth values.  For F13
electron beam flux densities, our approximations predict that smaller Balmer jump ratios and hotter blue-to-red optical
continua are produced in the models of dMe flares compared to flares
in the Sun and rapidly rotating dG stars.  However, the lower gas density
initially below
the flare transition region in lower surface gravity stars causes a
larger downflow velocity in the CC \citep{Fisher1989}. A
much larger downflow velocity ($v$) and a smaller preflare gas density ($\rho$)
produces a comparable mass flux density (in units of $g$ cm$^{-2}$ s$^{-1}$) and thus a similar value of
$\rho_{\rm{maxCC}}$ and continuum
optical depth compared to the response of a higher gravity star to
an F13 beam flux density.  We discussed a new solar F13 flare model,
which may explain the radiation from superflares
in rapidly rotating dG stars; the density stratification template from this RADYN
model exhibits a much different velocity field than the CCs in the solar 5F11 and
dMe F13 models and can be used to make more refined
predictions for the emergent spectral properties at large values of
$m_{\rm{ref}}$ (e.g., in Figures
\ref{fig:grid} - \ref{fig:FcolorR}) when observational constraints of
the continuum flux ratios during flares exist for these stars.  

Using the density stratification of the CC in the solar 5F11 model from \citet{Kowalski2017A}
as a template for our approximate model atmospheres, one
can predict the emergent $\lambda = 2826$ \AA\ NUV continuum intensity in IRIS data of
solar flares, for which Fe II and Mg II line profiles suggest that
chromospheric condensations are produced in the
impulsive phase.  A relatively low
beam energy flux density (F11) with values of $E_c$ and $\delta$  that are
within the hard X-ray observational constraints but outside the limited parameter
space explored previously in the literature
\citep[e.g.,][]{Kowalski2017A}, does not reproduce the two brightest observed excess
IRIS NUV continuum intensity values in the 2014-March-29 X1 solar flare.  We have
confirmed this with a new RADYN calculation.   Therefore, a higher
flux density is required to reach the largest observed
$\lambda=2826$ \AA\ NUV continuum
brightness in this X-class flare, and our approximate calculations can
be used to guide RADYN modeling of other flares observed with IRIS.

In Appendix B, we present an extension of our model prescriptions:
calculations of the
emergent continuum flux from CCs with $T>>10,000$ K,
which was suggested to explain the optical continuum flux ratios in
\citet{Kowalski2012}. In future work, we will also extend our approximate prescriptions 
to RADYN atmospheres heated by electron beams with a high low-energy
cutoff $\gtrsim100$ keV.   Lower beam flux densities with a
high low-energy cutoff in the electron distribution can reproduce a 
small Balmer jump ratio, a hot red-to-blue optical continuum, and
narrow Balmer lines \citep[e.g., the $E_c=150$ keV 5F12 model in Appendix A
of][]{Kowalski2017B}.  

A Python GUI is freely available for the approximate evolved-1 and evolved-2 model
atmosphere calculations upon request to the first author.   
Appendix B demonstrates how broad wavelength continuum flux spectra can be constructed from our algorithm to fill in continuum regions of the flare
spectrum without constraints from observations.  If the Balmer jump ratio
of a flare is constrained, we suggest that our approximate
model atmospheres could be useful for modeling
the effects of ultraviolet flare radiation on exoplanet atmospheres
\citep[e.g.][]{Segura2010, Ranjan2017}. Approximations for
wavelength regimes that are not possible to observe in solar flares
will also be useful for addressing flare energy budget problems
\citep{Fletcher2007, Milligan2014, Kleint2016}.

\section*{Appendix A:  The Algorithm for the Temperature
  Stratification of our Approximate Flare Atmospheres}

In this Appendix, we present the details of our algorithm for
constructing a temperature
stratification for the approximate model atmospheres in the CC and in
the stationary flare layers.  

\subsection*{A.1. Temperature within the CC}
At the column masses in the CC that are less than where $T_{\rm{min CC}}$ occurs, we approximate the temperature from $T_{\rm{min CC}}$ to $T=17,000$ K as an exponential rise vs.  log $m$.  From $T=17,000$ K to $25,000$ K (at $x=0$), a linear rise with log $m$ is used.  The column mass corresponding to $T=25,000$ K is chosen as log $m_{\rm{ref}}-1.0$, since $\sim$10\% of the column mass is at higher temperatures in the flare transition region and corona.   The temperature stratification at low heights in the CC is assumed to be constant and equal to $T_{\rm{minCC}}$.  
In Figure \ref{fig:5F11CC}, there is a local temperature maximum at the highest column mass of the
CC in the RADYN calculation due to viscous and compressive heating
contributions \citep{Kowalski2015}; ignoring this feature in our
approximations does not make a difference in our
results.

\subsection*{A.2. Temperature within the stationary flare layers}

At the evolved times in the RADYN calculations, the temperature at the top of the stationary flare layers is set to the value of $T_{\rm{minCC}}$, which is determined from $T_{\rm{ref}}$ with the simple adjustments described in Section \ref{sec:method_param}. 
For harder electron beam ($\delta \sim 3$) simulations, the electron
density in the stationary flare layers is higher than in softer
($\delta \gtrsim 4$) beams.  We find that a linear temperature
decrease of 3000 K adequately reproduces the ambient electron density
in the \dmehard\ simulation and a linear temperature decrease of 5000
K adequately reproduces the ambient electron density in the \dmesoft\
and \solsoft\ simulations, which have softer distributions of
nonthermal electrons at the energies that heat the stationary flare
layers. The gravitational acceleration determines the height range
below the CC over which these temperature drops occur.  Due to the
differences in log $g$, the total physical depth range of the
stationary flare layers is $\Delta z=150$ km for the dMe and $\Delta z
=300$ km for the solar atmosphere.  

In summary, the temperature stratification of the stationary flare layers heated by electron beams are determined as follows:  

\begin{itemize}
\item Evolved-A, lower beam flux densities (5F11):  $T=T_{\rm{ref}}+500$ K at the top of the stationary flare layers and bottom of the CC ($x=x_{\rm{maxCC}}$); the temperature stratification linearly decreases by $\Delta T=3000$ K for hard electron beams ($\Delta T=5000$ K for soft electron beams) extending to $x=x_{\rm{maxCC}}+100$ km below the CC (for the dMe) and 200 km below the CC (for the solar atmosphere) .  

\item Evolved-2, lower beam flux densities (5F11):   $T=T_{\rm{ref}}$ at the top of the stationary flare layers and bottom of the CC ($x=x_{\rm{maxCC}}$); the temperature stratification linearly decreases by $\Delta T=3000$ K for hard electron beams ($\Delta T=5000$ K for soft electron beams) extending to $x=x_{\rm{maxCC}}+100$ km below the CC (for the dMe) and 200 km below the CC (for the solar atmosphere).  

\item Evolved-A, high beam flux densities (F13):  $T=T_{\rm{ref}}+1500$ K at the top of the stationary flare layers and bottom of the CC ($x=x_{\rm{maxCC}}$); the temperature stratification linearly decreases by $\Delta T=3000$ K for hard electron beams ($\Delta T=5000$ K for soft electron beams) extending to $x=x_{\rm{maxCC}}+100$ km below the CC (for the dMe) and 200 km below the CC (for the solar atmosphere).  

\item Evolved-2, high beam flux densities (F13):  $T=T_{\rm{ref}}+1000$ K at the top of the stationary flare layers and bottom of the CC ($x=x_{\rm{maxCC}}$); the temperature stratification linearly decreases by $\Delta T=3000$ K for hard electron beams ($\Delta T=5000$ K for soft electron beams) extending to $x=x_{\rm{maxCC}}+100$ km below the CC (for the dMe) and 200 km below the CC (for the solar atmosphere).  

\end{itemize}

For all approximate model atmospheres, the bottom of the stationary
flare layers extends to $x=x_{\rm{maxCC}}+150$ km for the dMe
($x=x_{\rm{maxCC}}+300$ km for the solar atmosphere).   We adjust the
physical depth range linearly 
for other values of the surface gravity, but no change is made to the
density in the stationary flare layers compared to the solar and dMe
pre-flare density stratification.  
 The temperature at the bottom of the stationary flare layers is set to $T=6500$ K.  At these lowermost flare layers in our approximate evolved models, the optical depth is either too large for any emission to escape or the emissivity is small compared to the emissivity in the higher layers.  Therefore, our results are not sensitive to the details at such low temperature.

\section*{Appendix B:  Superhot $T\sim10^5$ K CCs in \MakeLowercase{d}M\MakeLowercase{e} Flares?}

\begin{figure}[h!]
\plotone{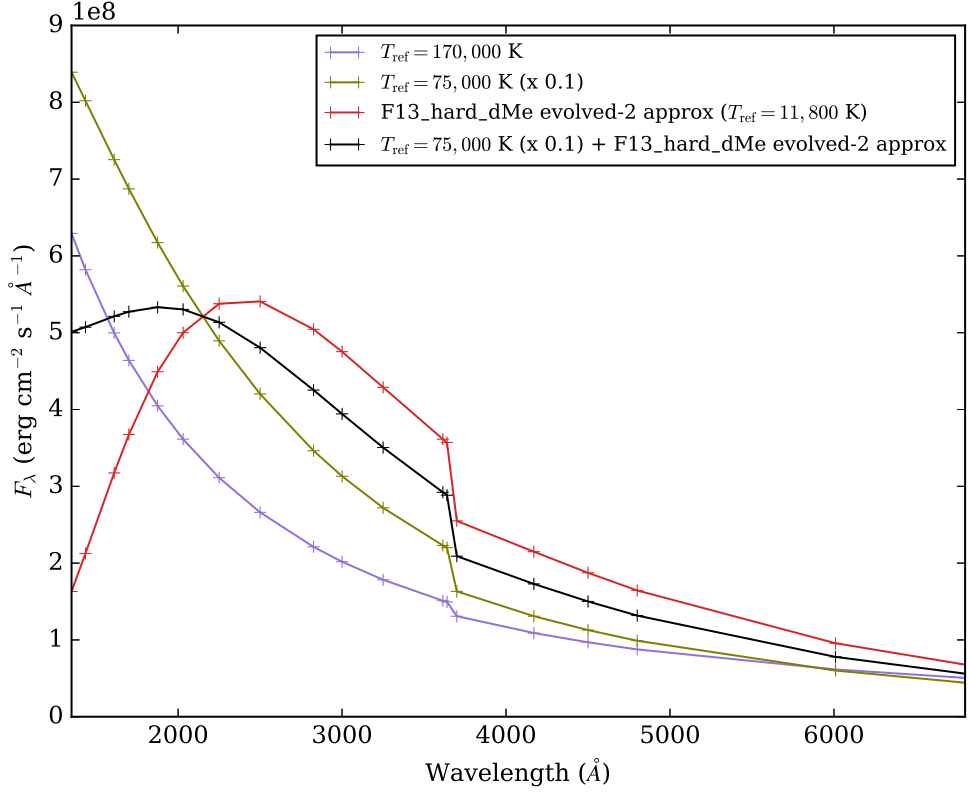}
\caption{ Flare spectra at $\lambda=1300$\AA\ to 6800\AA\ from approximate
  evolved atmosphere calculations that are all generally consistent with
  impulsive phase observations of dMe flares at $\lambda > 3500$ \AA.  
  Thermal bremsstrahlung radiation from a dense, superhot CC with
  $T=170,000$ K (log $m_{\rm{ref}}=-2.3$) is shown as the purple curve,
 a superhot CC with $T=75,000$ K (log $m_{\rm{ref}}=-1.7$) multiplied by
 0.1 is shown as the olive curve, the
  approximate evolved-2 \dmehard\ is shown as the red curve, and a
  superposition of the  $T=75,000$ K (log $m_{\rm{ref}}=-1.7$) CC model
  multiplied by 0.1 and the approximate evolved-2 \dmehard\ model is shown as
  the black curve.  For the superhot models, the emissivity from the
  stationary flare layers was not included in the calculation of the
  emergent radiative flux.  All models exhibit small Balmer jump
ratios, a color temperature at blue and red optical wavelengths of
$T_{\rm{FcolorR}}>8500$ K, and a rising spectrum from 3600 \AA\ to
2600 \AA.  The red spectrum peaks at $\lambda \sim
2500$ \AA\ and turns over into the FUV.  The superposed spectrum
exhibits a relatively flat distribution in the FUV and a peak 
near $\lambda \sim 2000$ \AA.  Spectral data at $\lambda < 2500$ \AA\ would be
able to constrain the role of very hot, very dense chromospheric
condensations in the impulsive phase of dMe flares.  The light purple
model exhibits a smaller Balmer jump ratio and FcolorR value compared to the 170,000 K slab model in
\citet{Kowalski2012} due to including the values of the Gaunt factors
in these new predictions.  
 }
  \label{fig:superhot}
\end{figure}

A ``superhot'' plasma with $T\sim170,000$ K produces the $\lambda>3500$
\AA\ continuum properties
that are generally consistent with some spectra in the impulsive phase
of a large YZ CMi flare with a red continuum exhibiting a cooler color temperature
\citep[cf Figure 7.6 of][and \citet{Kowalski2013}]{Kowalski2012}.  We calculate the emergent
spectrum from a ``superhot CC''  (where superhot refers to $T\sim10^5$ K
or greater) using our
evolved atmosphere approximations with $T_{\rm{minCC}}=170,000$ K and log
$m_{\rm{ref}}=-2.3$.  The flux spectrum is shown in Figure 
\ref{fig:superhot} from the far-ultraviolet through the optical\footnote{The
  dissolved level continuum opacity longward of the bound-free edges of
  hydrogen is not included in the spectra in Figure
  \ref{fig:superhot}.  We have included the continuum opacity from
  dissolved levels in our approximations and will discuss this in a
  future work.  }, which
exhibits a value of FcolorB $=1.5$ and FcolorR $=1.8$.  The evolved-1 density
in this CC is $n_e>10^{15}$ cm$^{-3}$ and results in a low
continuum optical depth $\tau_{\lambda} \lesssim 0.01$. The dominant 
continuum emissivity is free-free (thermal bremsstrahlung)
emission. The
contribution to the emergent intensity from the stationary flare layers is not included in this
calculation (the emissivity from these layers is set to 0) in order to
isolate the spectral properties of the superhot layers.  If stationary
flare layers at $T\sim10^4$ K are included in the calculation, then
the emergent spectrum exhibits a larger Balmer jump ratio due to the hydrogen
recombination radiation from the stationary flare layers with a comparable
electron density of $n_e \sim 10^{15}$ cm$^{-3}$. The superhot CC model exhibits a
physical depth range that is 
$10^4$ times larger than the path length (of several meters) of
material at $T=10^5-2 \times 10^5$ K
in the high flux density (5F11-F13) electron beam RADYN models.
The physical depth range of the $T\sim10^5$ K thermal bremsstrahlung would have to
be much larger than currently predicted in the RADYN models to
contribute significantly to the emergent radiation if there are also
stationary flare layers at lower temperature, as concluded for solar
flare spectra early on \citep{Hiei1982}.  

In the  $T=170,000$ K slab model in
\citet{Kowalski2012}, the Gaunt factors were assumed to be equal to 1, 
which results in a Balmer jump ratio and FcolorR values that
appear to be consistent with the observational comparison.  Including the
wavelength dependent Gaunt factors in the calculations here changes the spectral shape considerably.
We use the free-free Gaunt factors from \citep{Menzel1935} (as used in the RADYN
code) and bound-free Gaunt
factors from \citep{Seaton1960} (as used in the RH code) and find that
a very large value of log 
$m_{\rm{ref}}=-1.7$ and lower $T_{\rm{ref}} \sim 75,000$ K give more
reasonable values of the Balmer jump ratio ($\sim1.8$) and FcolorR values ($\sim2$) compared
to the observations in \citet{Kowalski2012}.  This superhot CC spectrum is shown in Figure
\ref{fig:superhot} scaled by a factor of 0.1.  Such a large value of $m_{\rm{ref}}$ produces $n_e
\sim 10^{16}$ cm$^{-3}$ in the CC. 

In Figure \ref{fig:superhot}
we also show the spectrum from the approximate \dmehard\
model at the evolved-2 time (with parameters given in the last row of Table
\ref{table:ccparams}), which produces heating to $T\sim10,000$ K at densities of $n>10^{15}$
cm$^{-3}$.  Compared to the superhot approximation $T_{\rm{ref}} \sim
75,000$ K, the \dmehard\ approximation exhibits a similar
color temperature at $\lambda>4000$ \AA\ and Balmer jump ratio but an opposite
continuum slope at $\lambda < 2500$ \AA.  
In the very early phases of the
\dmehard\ RADYN model ($t \sim 0.1$~s), the temperature of the CC over a
significant path length ($\Delta z\sim15$ km) is at a temperature of
$T\sim75,000$ K but this material has a much lower column mass than
log $m_{\rm{ref}} =
-1.7$.  Most of the emergent optical continuum radiation originates
from lower temperature at this very early time in the RADYN simulation, and the CC quickly cools below 75,000 K in a short
time ($\Delta t \lesssim 0.05$~s).

Too few spectral observations of the continuum at $\lambda<3500$ \AA\
exist during dMe flares to definitely rule out the existence of a
$T>50,000$ K CC in dMe flares.  Course
broadband photometry of a moderate-sized flare in AD Leo indicates that the spectral energy distribution
decreases from NUV to FUV wavelengths \citep{Hawley2003}.
An IUE/FUV spectrum exists for the first 900~s of the impulsive phase of the Great
Flare in AD Leo \citep{HP91}, showing a rather flat FUV continuum
distribution.  This flare exhibits a very low Balmer jump ratio, and 
\citet{HF92} demonstrates that the broadband photometry distribution rules
out a $T\sim1$ MK free-free continuum in favor of a
$T\sim9500$ K blackbody distribution.  NUV spectra at $\lambda<3500$ \AA\
were not obtained during the impulsive phase of this flare.  Clearly, more
NUV data at $\lambda<3500$ \AA\ are critical to completely rule out the rising spectral
characteristics of a superhot CC spectrum in some flares.  If we add
the \dmehard\ approximate evolved-2 spectrum
and the superhot ($T=75,000$ K) CC spectrum in Figure \ref{fig:superhot} we
obtain a nearly flat FUV continuum distribution as observed in the
Great Flare impulsive phase \citep[cf Figure 6 of][]{HP91}.  The
superhot CC spectrum was multiplied by a filling factor of 0.1
relative to the F13 approximate spectrum.   One
could speculate that a very short persistence of flare kernels with 
dense, $T=50,000-100,000$ K plasma and F13 beam-heated kernels (with an order
of magnitude larger area) may also help explain the faster time FUV time-evolution compared to the NUV \citep{Hawley2003} and the very high 
observed FUV/NUV flux ratios
in GALEX data \citep{Robinson2005, Welsh2006} of the impulsive phase of dMe flares.

\clearpage

\acknowledgements
We thank the referee Dr. P. Heinzel for his critique and comments
which significantly improved this work.  
AFK thanks Dr. M. Carlsson for discussions that led to the development of the ideas in this paper and for a critical reading of the
manuscript.  

\bibliography{ccmodel.bib}

\end{document}